\documentclass[a4paper,journal]{IEEEtran}
\usepackage[utf8]{inputenc}
\usepackage{amsmath,amssymb,amsfonts,amsthm,bm}
\usepackage{algorithmic}
\usepackage{algorithm}
\usepackage{array}
\usepackage{textcomp}
\usepackage{url}
\usepackage{verbatim}
\usepackage{graphicx}
\usepackage{cite}
\usepackage[caption=false,font=footnotesize]{subfig}

\usepackage{color}
\usepackage[acronyms]{glossaries}

\usepackage{threeparttablex}

\usepackage{pgfplots}
\pgfplotsset{compat=newest}
\usepgfplotslibrary{groupplots} 
\usepackage{tikz}
\usetikzlibrary{calc}
\usetikzlibrary{fit}
\usetikzlibrary{positioning}
\usetikzlibrary{spy}
\usetikzlibrary{arrows.meta}

\usetikzlibrary{external}
\usepackage[hidelinks]{hyperref}

\newcommand{\transpose}{\text{T}}
\newcommand{\hermitian}{\text{H}}

\newcommand{\diag}{\text{diag}}

\title{Fast Variational Block-Sparse Bayesian Learning
	\thanks{
		This research was partly funded by the Austrian Research PromotionAgency (FFG) within the project SEAMAL Front (project number: 880598). Furthermore, the financial support by the Christian Doppler Research Association, the Austrian Federal Ministry for Digital and Economic Affairs and the National Foundation for Research, Technology and Development is gratefully acknowledged.
		
		Jakob M\"oderl, Erik Leitinger and Klaus Witrisal are with the Institute of Communication Networks and Satellite Communications at Graz University of Technology, Graz Austria.
		
		Franz Pernkopf is with the Signal Processing and Speech Laboratory at Graz University of Technology, Graz, Austria.
		
		Bernard H. Fleury is with the Institute of Telecommunications, TU Wien, Vienna, Austria.
		
		Klaus Witrisal and Erik Leitinger are further associated with the Christian Doppler Laboratory for Location-aware Electronic Systems.
	}%
}

\author{
\IEEEauthorblockN{Jakob M\"oderl, Erik Leitinger, Bernard H. Fleury, Franz Pernkopf, and Klaus Witrisal} \\
}

\newtheorem{theorem}{Theorem}
\newtheorem{lemma}{Lemma}
\newtheorem{corollary}{Corollary}

\newcommand{\ist}{\hspace*{.3mm}}
\newcommand{\rmv}{\hspace*{-.3mm}}
\newcommand{\iist}{\hspace*{1mm}}

\newcommand{\nn}{\nonumber}

\newacronym{sbl}{SBL}{sparse Bayesian learning}
\newacronym{bsbl}{BSBL}{block-sparse Bayesian learning}
\newacronym{fsbl}{F-SBL}{fast-SBL}
\newacronym{fbsbl}{F-BSBL}{fast-BSBL}
\newacronym{vabsbl}{VA-BSBL}{variational BSBL}
\newacronym{vb}{VB}{variational Bayesian}
\newacronym{ml}{ML}{maximum likelihood}
\newacronym{pdf}{PDF}{probability density function}
\newacronym{snr}{SNR}{signal-to-noise ratio}
\newacronym{em}{EM}{expectation-maximization}
\newacronym{awgn}{AWGN}{additive white Gaussian noise}
\newacronym{nmse}{NMSE}{normalized mean squared error}
\newacronym{mmse}{MMSE} {minimum mean squared error}
\newacronym{elbo}{ELBO}{evidence lower bound}
\newacronym{kl}{KL}{Kullback-Leibler}
\newacronym{doa}{DOA}{direction of arrival}
\newacronym{mmv}{MMV}{multiple measurement vector}
\newacronym{ospa}{OSPA}{optimal subpattern assignment}
\newacronym{ar}{AR}{autoregressive}

\newlength{\figurewidth}
\newlength{\figureheight}

\begin{document}
	\maketitle
	\begin{abstract}
    We propose a \gls{vb} implementation of \gls{bsbl} 
    to compute proxy \glspl{pdf} that approximate the posterior \glspl{pdf} of the weights and associated hyperparameters in a block-sparse linear model, resulting in an iterative algorithm coined \gls{vabsbl}. The priors of the hyperparameters are selected to belong to the family of generalized inverse Gaussian distributions. 
    This family contains as special cases commonly used hyperpriors such as the Gamma and inverse Gamma distributions, as well as Jeffrey’s improper distribution.

    Inspired by previous work on classical \gls{sbl}, we investigate the update stage in which the proxy \glspl{pdf} of a single block of weights and of its associated hyperparameter are successively updated, while keeping the proxy \glspl{pdf} of the other parameters fixed. This stage defines a nonlinear first-order recurrence relation for the mean of the proxy \gls{pdf} of the hyperparameter. 
    By iterating this relation ``ad infinitum'' we obtain a criterion that determines whether the so-generated sequence of hyperparameter means converges or diverges. 
    Incorporating this criterion into the \gls{vabsbl} algorithm yields a fast implementation, coined \gls{fbsbl}, which achieves a two-order-of-magnitude runtime improvement. 
    
    We further identify the range of the parameters of the generalized inverse Gaussian distribution which result in an inherent pruning procedure that switches off ``weak'' components in the model, which is necessary to obtain sparse results.
    Lastly, we show that \gls{em}-based and \gls{vb}-based implementations of \gls{bsbl} are identical methods. Thus, we 
    extend a well-known result from classical \gls{sbl} to \gls{bsbl}.
    Consequently, \gls{fbsbl} and \gls{bsbl} using coordinate ascent to maximize the marginal likelihood coincide. These results provide a unified framework for interpreting existing \gls{bsbl} methods.
    
\end{abstract}
\glsresetall

\section{Introduction}
Sparse signal reconstruction has gained widespread adoption over the last 20 years through the advent of compressed sensing \cite{duarte2011TSP:CS}.
Generally, the aim of sparse signal reconstruction algorithms is to reconstruct a signal as a weighted linear combination of only a few entries from a large-size dictionary.
One approach to solve this problem is \gls{sbl}.
\Gls{sbl} uses a Gaussian scale mixture model \cite{barndorff-nielsen1982SR:normal-variance-mixture} in which the precision, i.e., the inverse variance, of each weight is interpreted as a hyperparameter.
	\footnote{Note that \gls{sbl} can be derived equivalently by using a hierarchical model on either the prior variances or the prior precisions of the weights.}
These hyperparameters are estimated from the data in a Type-II Bayesian fashion \cite{wipf2011TIP}, e.g., by directly maximizing the marginal likelihood \cite{tipping2001JMLR:SBL}, or using the \gls{em} algorithm for that purpose \cite{wipf2004TSP}, or by applying  \gls{vb} inference \cite{bishopUAI2000:variational-RVM}.
A major shortcoming of \gls{sbl} is its slow convergence.
In \cite{faul2001, tipping2003WAIS:FastMarginalSparseBayesian} a ``fast'' method using coordinate ascent to maximize the marginal likelihood is shown to alleviate this problem. 
Fast schemes have also been developed for \gls{em}- and \gls{vb}-based implementations of \gls{sbl} \cite{pedersen2015SP,shutin2011TSP:fastVSBL}. In both cases, fast update rules are obtained via the fixed points of a recurrent relation for the updates of a single hyperparameter.
	
In classical \gls{sbl}, all weights in the model are assumed independent.
However, some applications result in a block-sparse model in which groups or blocks of weights are jointly zero or nonzero.
Naturally, the \gls{sbl} approach, and its implementations, have been adapted to handle such block structures, an extension referred to as \gls{bsbl}.
For example, \cite{luessi2013TSP:VariationalBSBL} extends the marginal likelihood-based approach to a block structure.
Additionally, the authors show that the hyperparameters can be analytically determined by solving for the roots of a specific polynomial.
However, for the application outlined in this paper, the degree of this polynomial is considerably large. Therefore, solving for these roots is computationally inefficient and an iterative procedure is used instead.
In \cite{ma2017TSP:FastBSBL}, a fast \gls{bsbl} algorithm is presented based on the same polynomial solution as in \cite{luessi2013TSP:VariationalBSBL}.
Other \gls{bsbl} implementations rely on the \gls{em}-algorithm \cite{zhang2011STSP:bSBL,zhang2013TSP:BlockSparseSBL} or on \gls{vb} inference \cite{babacan2014TSP:variationalBSBL,sharma2018ICPR:VariationalBSBL}.
However, these \gls{bsbl} implementations suffer from slow convergence too.
	
The \gls{em}- and \gls{vb}-based implementations of classical \gls{sbl} are shown to be identical \cite{palmerNIPS2005}. This equivalence is conceptually relevant as the latter approach can be given a message-passing interpretation, in which messages are proxy \glspl{pdf}. This interpretation allows for a principled merging of classical \gls{sbl} and belief propagation \cite{riegler13}, e.g., for joint channel estimation and decoding \cite{hansenTSP2018:IterativeReceiverSBL}.
Additionally, different hyperpriors, i.e., priors on the hyperparameters, lead to different estimators.
By choosing a parameterized hyperprior \gls{pdf} that encompasses many commonly used hyperprior \glspl{pdf} as special cases, we can compare different variants of \gls{sbl} within the same framework \cite{babacan2014TSP:variationalBSBL,pedersen2015SP}.
A fast update rule for the hyperparameters is of critical importance for this analysis.
However, extending the fast variational \gls{sbl} algorithm presented in \cite{shutin2011TSP:fastVSBL} to block-sparse models is not trivial, since \cite[Theorem 1]{shutin2011TSP:fastVSBL} merely provides the condition for the existence of one (locally stable) fixed point, whereas multiple fixed points might exist in the block-sparse case.

\subsection*{Contribution}

This work presents three novel theoretical results regarding \gls{bsbl}.
\begin{itemize}
\item We first present a \gls{vabsbl} 
algorithm and then derive a fast implementation of it called \gls{fbsbl} algorithm.
To do so, we generalize \cite[Theorem 1]{shutin2011TSP:fastVSBL} to block-sparse models.

\item We use a generalized inverse Gaussian prior for the \gls{bsbl} hyperparameters. This distribution accounts for many commonly used hyperpriors, e.g., the Gamma distribution, inverse Gamma distribution and Jeffrey's improper distribution. Through the recurrent relation at the core of the \gls{fbsbl} algorithm, we are able to analyze which parameter settings of the generalized inverse Gaussian hyperprior results in an estimator incorporating an inherent pruning procedure that switches off ``weak'' components in the model, which is necessary to obtain sparse results.

\item We show that the \gls{em}- and \gls{vb}-based implementations of \gls{bsbl} are identical.
This result generalizes the corresponding equivalence shown for classical \gls{sbl} \cite{palmerNIPS2005} to the block-sparse case.
It further implies that \gls{fbsbl} and \gls{bsbl} using coordinate ascent to maximize the marginal likelihood, e.g., \cite{ma2017TSP:FastBSBL}, are identical.
Therefore, this result allows for a \gls{vb} message-passing interpretation of \gls{bsbl} algorithms, e.g., \cite{ma2017TSP:FastBSBL} or \cite{zhang2013TSP:BlockSparseSBL}.
\end{itemize}

In addition to the aforementioned theoretical results, we numerically demonstrate the following advantages of the proposed \gls{fbsbl} algorithm.
\begin{itemize}
\item The runtime of the \gls{fbsbl} algorithm is up to two orders of magnitude shorter than that of \gls{vabsbl} and related algorithms, e.g., the BSBL-EM algorithm (which is identical to a special case of \gls{vabsbl}) and the BSBL-BO algorithm, both of which are presented in \cite{zhang2013TSP:BlockSparseSBL}, while simultaneously achieving better performance in terms of \gls{nmse} and support recovery rate.
\item We apply the \gls{fbsbl} algorithm to a \gls{doa} estimation problem and show the benefits compared to the SBL-based \gls{doa} estimation algorithm for multiple measurement scenarios proposed in \cite{gerstoftSPL2016:SBLforDoA}.
\end{itemize}

\section{Overview of the BSBL Framework}
\label{sec:signal-model}
\subsection{System Model}
\label{sec:signal-model:setup}

We aim to estimate the weights $\bm{x}$ in the linear model
\begin{equation} \label{eq:signal-model}
\bm{y} = \bm{\Phi}\bm{x}+ \bm{v}
\end{equation}
where $\bm{y}$ is the observed signal vector of length $N$, $\bm{\Phi}$ is an $N\times M$ dictionary matrix and $\bm{v}$ is \gls{awgn} with precision $\lambda \in \mathbb{R}_{++}$, where $\mathbb{R}_{++} = \{x \in \mathbb{R}: x > 0\}$. 
Thus, the likelihood function is given by $p(\bm{y}|\bm{x},\lambda) = \mathrm{N}(\bm{y};\bm{\Phi}\bm{x},\lambda^{-1}\bm{I})$.
Here, $\mathrm{N}(\bm{x};\bm{\mu},\bm{\Sigma})=|\frac{\pi}{\rho}\bm{\Sigma}|^{-\rho} \exp\{-\rho(\bm{x}-\bm{\mu})^\hermitian\bm{\Sigma}^{-1}(\bm{x}-\bm{\mu})\}$, with $|\cdot|$ being the matrix determinant, denotes a multivariate Gaussian \acrshort{pdf} of the variable $\bm{x}$ with mean $\bm{\mu}$ and covariance matrix $\bm{\Sigma}$.
To allow for both a real-valued and complex-valued signal model, we use a likelihood function that is parameterized by $\rho=\frac{1}{2}$ for the real-valued case and $\rho=1$ for the complex-valued case.

We assume that $\bm{x}$ is block-sparse, meaning that the length-$M$ vector $\bm{x}$ is partitioned into $K$ blocks $\bm{x}_i$, $i=1,\ldots,K$ of known size $d$ each
\footnote{To simplify the notation, we assume equal block sizes.  All results can be straightforwardly extended to the case of unequal block sizes as well.}
such that all elements within a block are simultaneously zero or nonzero (with probability one), e.g.,
\begin{equation}\label{eq:blocked-weights-setup}
\bm{x} = [\underbrace{x_{1}\iist x_{2}\,\cdots\,x_{d}}_{\bm{x}_1^\transpose=\bm{0}^\transpose}\,\underbrace{x_{d+1}\ist\cdots\ist x_{2d}}_{\bm{x}_2^\transpose\neq \bm{0}^\transpose}\,\,\cdots\,\underbrace{x_{M-d+1}\,\cdots\, x_{M}}_{\bm{x}_K^\transpose=\bm{0}^\transpose}]^\transpose
\end{equation}
and most of the $K$ blocks are zero.
The block-sparse model arises in many applications, e.g., the estimation of block-sparse channels in multiple-input--multiple-output communication systems \cite{barbu2016TVT:OFDM-BSBL,prasadTSP2015:MIMO-BlockSBL}, or \gls{doa} estimation in multiple-measurement scenarios \cite{gerstoftSPL2016:SBLforDoA}.
We provide an example of how to apply the system model in \eqref{eq:signal-model-MMV} to \gls{doa} estimation in such multiple measurement scenarios in Section \ref{sec:application-doa}.
In that case, the signal modeled reads
\begin{equation} \label{eq:signal-model-MMV}
\bm{Y} = \bm{\Psi} \bm{X} + \bm{V}
\end{equation}
where $\bm{Y} = [\bm{y}_1\iist \bm{y}_2\,\cdots\,\bm{y}_J]$ is a matrix of $J$ measurements $\bm{y}_j$, $j=1,\ldots,J$, $\bm{\Psi}$ is some dictionary matrix with corresponding weights $\bm{X}$, and $\bm{V}$ is a matrix of additive noise.
A common assumption is that the sparsity profile is the same for each column in $\bm{X}$, i.e., that all elements in each row of $\bm{X}$ are either jointly zero or nonzero.
Finding the row-sparse matrix $\bm{X}$ in \eqref{eq:signal-model-MMV} is equivalent to finding the block-sparse weight vector $\bm{x}$ in \eqref{eq:signal-model}, where
$\bm{y} = \text{vec}(\bm{Y}^\transpose)$, $\bm{x} = \text{vec}(\bm{X}^\transpose)$, $\bm{v}=\text{vec}(\bm{V}^\transpose)$, and $\bm{\Phi} = \bm{\Psi} \otimes \bm{I}_J$ is the Kronecker product of $\bm{\Psi}$ with the $J \times J$ identity matrix $\bm{I}_J$. Here, $\text{vec}(\bm{X})$ denotes the operation of stacking the columns of the $N \times M$ matrix $\bm{X}$ into an $NM\times 1$ column vector.

\subsection{BSBL Probabilistic Model}
\label{sec:signal-model:probabilistic}
\Gls{bsbl} solves the sparse signal reconstruction task of \eqref{eq:signal-model} through the method of automatic relevance determination \cite{tipping2001JMLR:SBL}. Following the \gls{sbl} framework \cite{faul2001,bishopUAI2000:variational-RVM, tipping2003WAIS:FastMarginalSparseBayesian, wipf2004TSP,pedersen2015SP,shutin2011TSP:fastVSBL,luessi2013TSP:VariationalBSBL,zhang2011STSP:bSBL,zhang2013TSP:BlockSparseSBL,babacan2014TSP:variationalBSBL,ma2017TSP:FastBSBL}, we use a Gaussian scale mixture model \cite{barndorff-nielsen1982SR:normal-variance-mixture}, i.e., each block weight $\bm{x}_i$, $i=1,\ldots,K$ is Gaussian distributed with
\gls{pdf}
\begin{equation}
p(\bm{x}_i|\gamma_{i}) = \mathrm{N}\big(\bm{x}_i;\bm{0},(\gamma_i\bm{D})^{-1}\big)
\end{equation}
where $\gamma_i \in \mathbb{R}_{++}$ is a scaling hyperparameter and $\bm{D}$ is a known matrix characterizing the intra-block correlation. The matrix satisfies $\mathrm{tr}\{\bm{D}\}=d$ with $\text{tr}\{\cdot\}$ denoting the trace operator.
Given the hyperparameter vector $\bm{\gamma}=[\gamma_1\,\cdots\,\gamma_K]^\transpose$, 
the blocks of weights $\bm{x}_i$, $i=1,\ldots,K$ are conditionally independent, i.e., $p(\bm{x}|\bm{\gamma})=\prod_{i=1}^{K}p(\bm{x}_i|\gamma_i)$ .
By estimating $\bm{\gamma}$ from the data, the relevance of each block is automatically determined. It was observed experimentally, that the hyperparameter estimate of many blocks 
diverges to infinity. The corresponding components are effectively pruned from the model as the computed posterior probability distributions of the corresponding weights concentrate to zero.

To perform (approximate) Bayesian inference, we assume the hyperparmeters to be independent, identically distributed according to a distribution with \gls{pdf} $p(\gamma)$, 
i.e., $p(\bm{\gamma})=\prod_{i=1}^{K}p(\gamma_i)$.
With the above assumptions
the distribution of the weights has \gls{pdf} $p(\bm{x}) = \prod_{i=1}^{K} p(\bm{x}_i)$, where the $K$ factors are of the form
\begin{equation}\label{eq:prior-weights-type-I}
p(\tilde{\bm{x}}) = \int p(\tilde{\bm{x}}|\gamma)p(\gamma)\, \text{d}\gamma.
\end{equation}
Hence, by specifying different hyperprior \glspl{pdf} $p(\gamma)$, we obtain different priors $p(\tilde{\bm{x}})$ and, thus, different estimators of the block weights. 
Commonly used hyperpriors include the Gamma and inverse Gamma distributions
as well as Jeffrey's improper distribution with density $p(\gamma)= \gamma^{-1}$ \cite{shutin2011TSP:fastVSBL,pedersen2015SP,babacan2014TSP:variationalBSBL}.
The  generalized inverse Gaussian distribution includes these distributions as special cases for particular settings of its parameters.
Furthermore, the generalized inverse Gaussian distribution is conjugate for the likelihood $\gamma\mapsto p(\tilde{\bm{x}}|\gamma)$, which allows for solving the \gls{vb} updates analytically \cite{babacan2014TSP:variationalBSBL}.
The \gls{pdf} of the generalized inverse Gaussian distribution is given by \cite{jorgensen1982}
\begin{equation} \label{eq:generalized-inverse-gaussian}
p(\gamma;a,b,c)=\frac{\big(\frac{a}{b}\big)^{\frac{c}{2}}\gamma^{c-1}}{2K_{c}(\sqrt{a b})} \exp\big\{-\frac{1}{2}(a\gamma + b \gamma^{-1})\big\}
\end{equation}
with $(a,b,c)\in\{\Theta_{c}\times\{c\} : c\in\mathbb{R}\}$ where
	\begin{equation} \label{eq:gig-prior-parameter-ranges}
		\Theta_{c} = \begin{cases}
			\{(a,b) : a>0,\ b\geq 0\} & \text{if } c>0 \\
			\{(a,b) : a > 0,\ b > 0\} & \text{if } c=0 \\
			\{(a,b) : a\geq 0,\ b > 0\} & \text{if } c<0
		\end{cases}
		\,.
	\end{equation}
Inserting \eqref{eq:generalized-inverse-gaussian} into \eqref{eq:prior-weights-type-I} and solving yields the \gls{pdf} of $\tilde{\bm{x}}$:
	\begin{align}
		\label{eq:ghd}
		p(\tilde{\bm{x}}) \propto \frac{K_{-(c+\rho d)}\Big(\sqrt{b(a+\tilde{\bm{x}}^{\hermitian}\bm{D}\tilde{\bm{x}})}\Big)}{\big(\sqrt{b(a+\tilde{\bm{x}}^{\hermitian}\bm{D}\tilde{\bm{x}})}\big)^{c+\rho d}}
	\end{align}
i.e., the \gls{pdf} of the generalized hyperbolic distribution \cite{babacan2014TSP:variationalBSBL}.

We will also consider densities of improper distribution, like Jeffrey's density, which results with the setting $a=b=c=0$, as long as the proxy densities of $\bm{\gamma}$ introduced in Section~\ref{sec:signal-model:variational-inference} are \glspl{pdf}. 
We do so by extending the parameter space $\{\Theta_{c}\times\{c\} : c\in\mathbb{R}\}$ to
\begin{equation}\label{eq:Theta}
    \Theta=\{(a,b,c)\in \mathbb{R}_{+}^{2}\times \mathbb{R}: b>0 \ \lor \ c>-\rho d\}
\end{equation} 
where $\lor$ denotes the logical ``or'' operator, and $\mathbb{R}_{+} = \{x\in\mathbb{R}:x\geq 0\}$.

To complete the Bayesian model, we also need a prior for the noise precision $\lambda$. We use a Gamma distribution with \gls{pdf}
\begin{equation} \label{eq:noise-prior}
p(\lambda;\epsilon, \eta) = \frac{\eta^{\epsilon}}{\Gamma(\epsilon)}\lambda^{\epsilon-1} \exp\{-\eta \lambda\}
\end{equation}
where $\epsilon \in \mathbb{R}_{++}$ and $\eta \in \mathbb{R}_{++}$ are, respectively, the shape and the rate parameters and 
$\Gamma(\cdot)$ denotes the Gamma function. The Gamma distribution is conjugate for the precision of a Gaussian distribution. As done for the hyperprior \gls{pdf}, we will also consider the case $\epsilon=\eta=0$, which yields Jeffrey's density.

\subsection{Variational Bayesian Inference}
\label{sec:signal-model:variational-inference}

We apply \gls{vb} inference \cite{tzikas2008:VAEM} \cite[Ch. 10]{Bishop2006} to approximate the posterior \gls{pdf}
\begin{equation}\label{eq:posterior-distribution}
p(\bm{x},\bm{\gamma},\lambda|\bm{y})\propto p(\bm{y}|\bm{x},\lambda)p(\bm{x}|\bm{\gamma})p(\bm{\gamma})p(\lambda)
\end{equation}
of the parameter tuple $(\bm{x},\bm{\gamma},\lambda)$ 
with a ``simpler'' proxy \gls{pdf} $q_{\bm{x},\bm{\gamma},\lambda}(\bm{x},\bm{\gamma},\lambda)$. Specifically, we apply the structured mean-field theory and postulate that $q_{\bm{x},\bm{\gamma},\lambda}$ factorizes according to 
\begin{equation} \label{eq:variational-approximation}
q_{\bm{x},\bm{\gamma},\lambda}(\bm{x},\bm{\gamma},\lambda)=q_{\bm{x}}(\bm{x})q_{\lambda}(\lambda)\prod_{i=1}^{K}q_{\gamma_i}(\gamma_i)
.
\end{equation}
We seek the (optimal) factors in \eqref{eq:variational-approximation} 
that maximize an \gls{elbo}
$\mathcal{L}(q_{\bm{x},\bm{\gamma},\lambda})\leq \ln p(\bm{y})$
\begin{align}\label{eq:elbo}
    \mathcal{L}(q_{\bm{x},\bm{\gamma},\lambda}) = \big< \ln p(\bm{x}, \bm{\gamma}, \lambda, \bm{y}) - \ln q_{\bm{x},\bm{\gamma},\lambda}(\bm{x},\bm{\gamma},\lambda)\big>_{q_{\bm{x},\bm{\gamma},\lambda}}
\end{align}
where the joint \gls{pdf} $p(\bm{x},\bm{\gamma},\lambda,\bm{y})$ is given by the right-hand expression of \eqref{eq:posterior-distribution}, and $\big<\cdot\big>_{q}$ denotes the expectation of the function given as argument in the brackets with respect to the probability distribution with \gls{pdf} $q$. Equivalently, these optimal factors in \eqref{eq:variational-approximation} minimize the \gls{kl}-divergence 
of the proxy \gls{pdf} $q_{\bm{x},\bm{\gamma},\lambda}$ from the true posterior \gls{pdf} $p(\bm{x},\bm{\gamma},\lambda|\bm{y})$ \cite[Ch. 10]{Bishop2006}.
In the sequel we denote these solutions with the superscript $\cdot^\star$, e.g., $q^\star_{\bm{x}}$.
For any $\iota \in \{\bm{x},\,\lambda,\,\gamma_1,\,\gamma_2,\,\cdots,\,\gamma_K\}$ the optimal factor $q^\star_{\iota}$ fulfills the consistency equation
    \footnote{In the following we use $q_{\iota}$ as a shorthand for $q_\iota(\iota)$ with $\iota \in \{\bm{x},\,\lambda,\,\gamma_1,\,\gamma_2,\,\cdots,\,\gamma_K\}$.}
\begin{equation} \label{eq:variational-update}
q^\star_\iota(\iota) \propto \exp\big\{\big<\ln p(\bm{x},\bm{\gamma},\lambda|\bm{y})\big>_{q_{\sim\iota}}\big\}
\end{equation}
where $q_{\sim\iota}$ denotes the product of all factors of $q$ in \eqref{eq:variational-approximation} but $q_\iota$.

\begin{table*}
\begin{threeparttable}
	\centering
	\caption{Fast update rules of $\hat{\gamma}_i$ for different parameterizations of $p(\gamma)$, expanding on \cite[Table 1]{babacan2014TSP:variationalBSBL}$^\ast$}
	\label{tab:update-equations}
	\begin{tabular}{ccccccc}
		\hline \hline \\[-2mm]
		& Hyperprior Parameters & \parbox{2.2cm}{\centering Hyperprior Density $p(\gamma)$} & Update of $\hat{\gamma}_i$  & $f_i(\gamma)$ & Fast Update Polynomial $G_i(\gamma)$ & \parbox{1.5cm}{\centering Pruning of Components} \\ \hline \\[-2mm]
		i & $(a,b,c) \in \Theta$ & $\gamma^{c-1}e^{-\frac{1}{2}(a\gamma+\frac{b}{\gamma})}$ &  $\sqrt{\frac{b}{\hat{a}_i}}  \frac{K_{\hat{c}+1} \big(\sqrt{\hat{a}_i\ist b} \big)}{K_{\hat{c}}\big(\sqrt{\hat{a}_i \ist b} \big)}$ & Eq. \eqref{eq:update-function-general-form} & - & \parbox{1.5cm}{\centering If both $a\rmv\rmv=\rmv\rmv 0$ and $c \rmv\geq\rmv 0$} \\ \\[-2mm]
		ii & $a \rmv = \rmv 0$, $b \rmv > \rmv 0$, $c \rmv = \rmv -\rho d - \frac{1}{2}$ & $\gamma^{c-1}e^{-\frac{b/2}{\gamma}}$  & $\sqrt{\frac{b}{2\rho\big<\bm{x}_i^\hermitian\bm{D}\bm{x}_i\big>_{q_{\bm{x}}}}}$  & $\sqrt{\frac{b A_i(\gamma)}{2\rho B_i(\gamma)}}$ & \scalebox{0.9}{$b A_i(\gamma) - 2\rho \gamma^{2} B_i(\gamma)$} & No \\ \\[-2.5mm]
		iii & $a > 0$, $b = 0$, $c>-\rho d$ & $\gamma^{c-1}e^{-\frac{a}{2}\gamma}$ & $\frac{c + \rho d}{\rho \big<\bm{x}_i^\hermitian\bm{D}\bm{x}_i\big>_{q_{\bm{x}}} + \frac{a}{2}}$ & $\frac{c+\rho d}{\rho\frac{B_i(\gamma)}{A_i(\gamma)}+\frac{a}{2}}$ & \scalebox{0.9}{$(c \rmv\rmv + \rmv\rmv \rho d ) A_i(\gamma) - \gamma \big(\rho B_i(\gamma)\rmv\rmv+\rmv\rmv \frac{a}{2} A_i(\gamma)\big)$} & No \\ \\[-3mm]
		iv  & $a = b = 0$, $c>-\rho d$ & $\gamma^{c-1}$ & $\frac{c + \rho d}{\rho \big<\bm{x}_i^\hermitian\bm{D}\bm{x}_i\big>_{q_{\bm{x}}}}$ & $\frac{(c + \rho d)A_i(\gamma)}{\rho B_i(\gamma)}$ & \scalebox{0.9}{$(c \rmv\rmv + \rmv\rmv \rho d) A_i(\gamma) - \rho \gamma B_i(\gamma)$} & If $c\geq 0$ \\ \\[-3mm]
		v & $a = b = c = 0$ & $\gamma^{-1}$ & $\frac{d}{\big<\bm{x}_i^\hermitian\bm{D}\bm{x}_i\big>_{q_{\bm{x}}}}$ & $\frac{d A_i(\gamma)}{B_i(\gamma)}$ & \scalebox{0.9}{$d A_i(\gamma) - \gamma B_i(\gamma)$} & Yes
		\\[3mm]
		\hline \hline
	\end{tabular}
	\begin{tablenotes}
        \item[$\ast$] When comparing the entries in this table and those in \cite[Table 1]{babacan2014TSP:variationalBSBL}, note that we use precision hyperparameters $\gamma_i$ while \cite{babacan2014TSP:variationalBSBL} uses variances $\gamma_i^{-1}$ instead.
	\end{tablenotes}
\end{threeparttable}
\end{table*}

\subsubsection*{Consistency equation for $q^\star_\lambda$}
For $\iota=\lambda$ in \eqref{eq:variational-update} we obtain
\begin{equation}\label{eq:noise-proxypdf}
q^\star_{\lambda}(\lambda) \propto \lambda^{\hat{\epsilon}-1} \exp\{-\hat{\eta} \lambda\}\, ,
\end{equation}
i.e., a Gamma \gls{pdf} with shape $\hat{\epsilon}=\rho N+\epsilon$ and rate $\hat{\eta}=\rho  \big(\|\bm{y}-\bm{\Phi}\hat{\bm{x}}\|^2 + \text{tr}(\bm{\Phi}^\hermitian\bm{\Phi}\hat{\bm{\Sigma}})\big)+\eta$ 
with $\hat{\bm{x}}$ and $\hat{\bm{\Sigma}}$ defined in \eqref{eq:weights-mean}, and $\|\cdot\|$ denoting the Euclidean norm. We will see that the other optimal factors depend on $q_\lambda$ only via its mean. For $q^\star_\lambda$ given in \eqref{eq:noise-proxypdf} we get
\begin{equation} \label{eq:lambda-update-rule}
\hat{\lambda} = \big<\lambda\big>_{q^\star_{\lambda}} =
\frac{\rho N+\epsilon}{\rho \big(\|\bm{y}-\bm{\Phi}\hat{\bm{x}}\|^2 + \text{tr}(\bm{\Phi}^\hermitian\bm{\Phi}\hat{\bm{\Sigma}})\big)+\eta}
\ist .
\end{equation}

\subsubsection*{Consistency equation for $q^\star_{\bm{x}}$}
For $\iota=\bm{x}$ in \eqref{eq:variational-update} we obtain
\begin{equation} \label{eq:weight-proxypdf}
    q^\star_{\bm{x}}(\bm{x}) = \mathrm{N}(\bm{x};\hat{\bm{x}},\,\hat{\bm{\Sigma}})
\end{equation}
with $\hat{\bm{x}}$ and $\hat{\bm{\Sigma}}$ given by
\begin{align}\label{eq:weights-mean}
\hat{\bm{x}} &= \hat{\lambda}\hat{\bm{\Sigma}}\bm{\Phi}^{\hermitian}\bm{y}
& &\text{and} &
\hat{\bm{\Sigma}} &= (\hat{\lambda}\bm{\Phi}^\hermitian\bm{\Phi} + \hat{\bm{\Gamma}})^{-1}\, ,
\end{align}
respectively, where $\hat{\bm{\Gamma}}=\diag(\hat{\bm{\gamma}})\otimes\bm{D}$, $\hat{\bm{\gamma}}=[\hat{\gamma}_1\ist\iist\hat{\gamma}_2\ist\cdots\ist\hat{\gamma}_K]^\transpose$ with $\hat{\gamma}_i$, 
$i=1,\dots,K$, given by \eqref{eq:expectation-gamma-inverse-general}, and $\diag(\cdot)$ denotes a square diagonal matrix with diagonal elements equal to the elements of the vector given as an argument.

\subsubsection*{Consistency equation for $q^\star_{\gamma_i}$} 
Finally, for $\iota=\gamma_i$, $i=1,\dots,K$ in \eqref{eq:variational-update} we show that
\begin{equation}\label{eq:precision-distribution}
q^\star_{\gamma_i}(\gamma_i) \propto \gamma_i^{\hat{c}-1} \exp\Big\{-\frac{1}{2}(\hat{a}_i\gamma_i + b \gamma_i^{-1})\Big\}\,,
\end{equation}
i.e., $q^\star_{\gamma_i}(\gamma_i)$ equals the \gls{pdf} of a generalized inverse Gaussian distribution \eqref{eq:generalized-inverse-gaussian} with parameters \cite{babacan2014TSP:variationalBSBL}
\begin{align} \label{eq:gig-param-ai-update}
\hat{a}_i&= 2\rho \big<\bm{x}_i^\hermitian\bm{D}\bm{x}_i\big>_{q_{\bm{x}}} \rmv\rmv + a
\, , & b& \, , & \hat{c} &= c+\rho d \,.
\end{align}
Note that $\hat{a}_i$ effectively only depends on the marginal \gls{pdf} of $\bm{x}_i$ obtained from $q_{\bm{x}}$.
From \eqref{eq:gig-param-ai-update} we find that $\hat{a}_i>0$. It follows from \eqref{eq:gig-prior-parameter-ranges} that each proxy PDF $q^\star_{\gamma_i}$ is a proper \gls{pdf} provided $(a,b,c)\in \Theta$, i.e., if $a\geq 0$ and either $b>0$ or $c>-\rho d$.
In this case, the mean of $q^\star_{\gamma_i}$ is computed to be \cite{babacan2014TSP:variationalBSBL, jorgensen1982}
\begin{align} \label{eq:expectation-gamma-inverse-general}
\hat{\gamma}_i &= \big<\gamma_i\big>_{q^\star_{\gamma_i}} = \sqrt{\frac{b}{\hat{a}_i}}  \frac{K_{\hat{c}+1} \big(\sqrt{\hat{a}_i b} \big)}{K_{\hat{c}}\big(\sqrt{\hat{a}_i b} \big)}
\ist .
\end{align}

Simplified expressions of the right-hand side in \eqref{eq:expectation-gamma-inverse-general} for particular selections of the parameters $a$, $b$ and $c$ are listed in the fourth column of Table \ref{tab:update-equations}. Note that the densities $p(\gamma)$ in Row~ii and Row~iii of Table~\ref{tab:update-equations} correspond to the inverse Gamma \gls{pdf} and the Gamma \gls{pdf}, respectively, when properly normalized. The density in Row~v is Jeffrey's density.

An iterative algorithm for computing the optimal factors
given in \eqref{eq:lambda-update-rule}, \eqref{eq:weights-mean}, and \eqref{eq:expectation-gamma-inverse-general} with $i=1,\ldots,K$
or, specifically, their parameters is obtained as follows: estimates of these parameters are first initialized and then successively updated in a round-robin fashion. The update steps are obtained by reinterpreting the equations in \eqref{eq:lambda-update-rule}, \eqref{eq:weights-mean}, and \eqref{eq:expectation-gamma-inverse-general}  with $i=1\ldots,K$, as updating rules as follows. A revised estimate of the left-hand parameter of each equation is computed using the right-hand expression with all occurring parameters replaced by their current estimate.

This iterative process is carried out until a convergence criterion is fulfilled or a maximum number of iterations is reached. We refer to this algorithm as \acrfull{vabsbl}.

\section{Variational Fast Solution}
\label{sec:variational-fast-solution}

Experimental evidence shows that the convergence of \gls{vabsbl} can be slow.
To derive a fast implementation of \gls{vabsbl}, which we coin \acrfull{fbsbl}, we follow the approach of \cite{shutin2011TSP:fastVSBL}.
Let us consider the stage consisting of first updating $q_{\bm{x}}$ using \eqref{eq:weight-proxypdf} and a current estimate 
of $\hat{\bm{\gamma}}$ followed by updating $q_{\gamma_i}$ using \eqref{eq:precision-distribution}.
The parameter $\hat{a}_i$ in \eqref{eq:gig-param-ai-update} with $q_{\bm{x}}=q_{\bm{x}}^\star$ given in \eqref{eq:weight-proxypdf} depends on the current estimate of $\hat{\gamma}_i$ through the expectation $\langle\bm{x}_i^{\hermitian}\bm{D}\bm{x}_i\rangle_{q_{\bm{x}}^\star}$. Taking the mean of $q^\star_{\gamma_i}$ yields the revised estimate of $\hat{\gamma}_i$. Thus, this stage computes a revised estimate of $\hat{\gamma}_i$ from the current estimate of $\hat{\gamma}_i$. 
We analyze the sequence of estimates $\{\hat{\gamma}_i^{[n]}\}_{n=0}^{\infty}$ generated by repeatedly executing this update stage ad infinitum, starting from an initial value $\hat{\gamma}_i^{[0]}$, while keeping the other proxy \glspl{pdf} fixed.

The sequence can be viewed as generated by a first-order recurrence relation which specifies for each sequence element the next element according to $\hat{\gamma}_i^{[n+1]}=f_i(\hat{\gamma}_i^{[n]})$ using the update function $f_i$ : $\mathbb{R}_{++} \mapsto \mathbb{R}_{++}$ obtained from \eqref{eq:expectation-gamma-inverse-general} to be
\begin{equation}\label{eq:update-function-general-form}
	f_i(\gamma) = \sqrt{\frac{b}{\hat{a}_i(\gamma)}}\frac{K_{\hat{c}+1}\big(\sqrt{b\, \hat{a}_i(\gamma)}\big)}{K_{\hat{c}}\big(\sqrt{b\, \hat{a}_i(\gamma)}\big)}
\end{equation}
with $\hat{a}_i(\gamma)$ denoting $\hat{a}_i$ in \eqref{eq:gig-param-ai-update} with $q_{\bm{x}}=q_{\bm{x}}^\star$ in \eqref{eq:weight-proxypdf} and $\hat{\gamma}_i$ in \eqref{eq:weights-mean} replaced by the variable $\gamma$.

We are interested in knowing whether the sequence $\{\hat{\gamma}_i^{[n]}\}_{n=0}^{\infty}$ converges or not and, if it does converge, in its limit.
If the sequence converges, the limit must be a fixed point $\gamma^\ast$ of the recurrence relation, i.e., it must fulfill
\begin{equation}\label{eq:fixed-point-equation}
f_i(\gamma^\ast)-\gamma^\ast=0
\, .
\end{equation}
The following Lemmas~\ref{lemma:expectation-as-polynomial-fraction} and \ref{lemma:decreasing-expectation} state that the expectation $\big<\bm{x}_i^\hermitian\bm{D}\bm{x}_i\big>_{q_{\bm{x}}^\star}$ is a strictly decreasing rational function of $\hat{\gamma}_i$. Lemma~\ref{lemma:besselK-ratio} and Corollary~\ref{corollaary:fi-increasing} state
that $f_i$ is strictly increasing.
Thus, Theorem~\ref{theorem:convergece}, a variant of the monotone convergence theorem \cite[Theorem 3.14]{Rudin1964}, can be used to determine under which condition the sequence $\{\hat{\gamma}_i^{[n]}\}_{n=1}^{\infty}$ converges or diverges based on the set of fixed points of the update function $f_i$.

\begin{lemma}
	\label{lemma:expectation-as-polynomial-fraction}
	The term $\big<\bm{x}_i^{\hermitian}\bm{D}\bm{x}_i\big>_{q_{\bm{x}}^\star}$ in \eqref{eq:gig-param-ai-update} can be expressed as a rational function of the current estimate $\hat{\gamma}_i$, i.e., $\big<\bm{x}_i^{\hermitian}\bm{D}\bm{x}_i\big>_{q_{\bm{x}}^\star}=B_i(\hat{\gamma}_i)/A_i(\hat{\gamma}_i)$ , where $A_i(\gamma)$ and $B_i(\gamma)$ are polynomials of degree $2d$ and $2d-1$, respectively, with positive coefficients. 
\end{lemma}	
\begin{proof}
	See Appendix~\ref{sec:variational-fast-solution:update-polynomials}.
\end{proof}
Note that by inserting $\big<\bm{x}_i^{\hermitian}\bm{D}\bm{x}_i\big>_{q_{\bm{x}}^\star} = B_i(\hat{\gamma}_i)/A_i(\hat{\gamma}_i)$ from Lemma~\ref{lemma:expectation-as-polynomial-fraction} into the simplified updates of $\hat{\gamma}_i$ given in the fourth column of Table~\ref{tab:update-equations}, we obtain simplified expressions for the update function $f_i$, which are given in the fifth column of this table.
Moreover, by inserting these simplified expressions for the update function $f_i$ into the fixed-point equation \eqref{eq:fixed-point-equation}, we find the fixed points $\gamma^\ast$ as the roots of polynomials $G_i(\gamma)$, which are given in the sixth column of Table~\ref{tab:update-equations}.

\begin{lemma}
	\label{lemma:decreasing-expectation}
	The rational function $B_i(\gamma)/A_i(\gamma)$ is strictly decreasing on its domain $\mathbb{R}_{++}$.
\end{lemma}
\begin{proof}
	See Appendix~\ref{sec:appendix:derivative-hi}.
\end{proof}

\begin{lemma}
	\label{lemma:besselK-ratio}
	The function $h_{\alpha}(u) = u^{-1/2} K_{\alpha+1}(\sqrt{u})/K_{\alpha}(\sqrt{u})$ defined on $\mathbb{R}_{++}$ is strictly decreasing for all $\alpha \in \mathbb{R}$.
\end{lemma}
\begin{proof}
	See Appendix~\ref{sec:appendix:fi-increasing}.
\end{proof}

\begin{corollary}
	\label{corollaary:fi-increasing}
	The update function $f_i$ given in \eqref{eq:update-function-general-form} is strictly increasing.
\end{corollary}
\begin{proof}
	Let's define $u_i(\gamma)=b\, \hat{a}_i(\gamma) = 2\rho b\, B_i(\gamma)/A_i(\gamma) + a\, b$, such that the right-hand expression in \eqref{eq:update-function-general-form} can be written as the composition
    $b\, h_{\hat{c}} \circ u_i(\gamma)$, where $h_{\hat{c}}(u)$ is defined in Lemma~\ref{lemma:besselK-ratio}.
	According to Lemmas~\ref{lemma:decreasing-expectation} and \ref{lemma:besselK-ratio} the functions $h_{\hat{c}}(u)$ and $u_i(\gamma)$ are strictly decreasing and, thus, the above composition is strictly increasing.
\end{proof}

\begin{theorem}
\label{theorem:convergece} Given any $i=1,\dots,K$, let $\mathcal{G}_i = \{ \gamma^\ast\in \mathbb{R}_{++} : f_i(\gamma^\ast) - \gamma^\ast = 0 \}$ be the set of fixed points of $f_i$ in \eqref{eq:update-function-general-form}. 
Then, the convergence of the sequence of estimates $\{\hat{\gamma}_i^{[n]}\}_{n=1}^{\infty}$ generated by the first-order recurrence with update function $f_i$ is determined by the initial value $\hat{\gamma}_i^{[0]}$ as follows:
\begin{equation} \label{eq:update-sequence-limit}
    \hspace*{-2ex}
	\lim_{n\rightarrow \infty} \hat{\gamma}_i^{[n]} = 
	\begin{cases}
		\displaystyle
		\infty & \mathrm{if}\ f_i(\hat{\gamma}_i^{[0]})>\hat{\gamma}_i^{[0]} \ \mathrm{and}\  \mathcal{G}_i^{+} = \emptyset\\
		\min \mathcal{G}_i^{+} & \mathrm{if}\ f_i(\hat{\gamma}_i^{[0]}) > \hat{\gamma}_i^{[0]} \ \mathrm{and}\  \mathcal{G}_i^{+} \neq \emptyset\\
		\max \mathcal{G}_i^{-} & \mathrm{if}\  f_i(\hat{\gamma}_i^{[0]}) \leq \hat{\gamma}_i^{[0]}\,
	\end{cases}
\end{equation}
where $\mathcal{G}_i^{+} = \mathcal{G}_i^{+}(\hat{\gamma}_i^{[0]})= \{\gamma^\ast \in \mathcal{G}_i\,:\,\gamma^\ast > \hat{\gamma}_i^{[0]}\}$, $\mathcal{G}_i^{-} = \mathcal{G}_i^{-}(\hat{\gamma}_i^{[0]})= \{\gamma^\ast \in \mathcal{G}_i\,:\,\gamma^\ast \leq \hat{\gamma}_i^{[0]}\}$, and $\emptyset$ is the empty set. 
\end{theorem}
Note that $\mathcal{G}_i^{-} \neq \emptyset$ if $f_i(\hat{\gamma}_i^{[0]})\leq\hat{\gamma}_i^{[0]}$, as shown in the following proof.
\begin{proof} 
Since $f_i$ is strictly increasing, the sequence $\{\hat{\gamma}_i^{[n]}\}_{n=1}^{\infty}$ is either strictly increasing if $f_i(\hat{\gamma}_i^{[0]})>\hat{\gamma}_i^{[0]}$ (Case~1), or strictly decreasing if $f_i(\hat{\gamma}_i^{[0]})<\hat{\gamma}_i^{[0]}$ (Case~2), while the case $f_i(\hat{\gamma}_i^{[0]})=\hat{\gamma}_i^{[0]}$ is trivial.
By definition, every fixed point $\gamma^\ast \in \mathcal{G}_i$ must fulfill $f_i(\gamma^\ast) = \gamma^\ast$. Assume first that $\mathcal{G}_i^{+}$ and $\mathcal{G}_i^{-}$ are non-empty. Since $f_i$ is strictly increasing, the sequence $\{\hat{\gamma}_i^{[n]}\}_{n=1}^{\infty}$ is bounded above by $\min \mathcal{G}_i^{+}$ in Case~1 and bounded below by $\max \mathcal{G}_i^{-}$ in Case~2. The monotone convergence theorem \cite[Theorem 3.14]{Rudin1964} states that the sequence actually converges to either of these values. In Case~2, the sequence cannot diverge to $-\infty$ because the range of $f_i$ is strictly positive. Thus, a fixed point $\gamma^\ast$ must exist in the interval $(0,\,\hat{\gamma}_i^{[0]}]$ and, consequently, $\mathcal{G}_i^{-}$ cannot be empty.
In Case~1, the condition $\mathcal{G}_i^{+}=\emptyset$ implies that there exists $\delta>0$ such that $f_i(\hat{\gamma}_i^{[n]})-\hat{\gamma}_i^{[n]}>\delta$ for infinitely many values of $n$. Thus, $\{\hat{\gamma}_i^{[n]}\}_{n=1}^{\infty}$ diverges to infinity.

\end{proof}

In case the fixed points in $\mathcal{G}_i$ can be found as roots of a polynomial, e.g., the special cases given in rows ii--v of Table~\ref{tab:update-equations}, we derive a fast update rule for $q_{\gamma_i}$ and $q_{\bm{x}}$ as follows. First, the fixed points in $\mathcal{G}_i$ are obtained by solving for the roots of the polynomial $G_i(\gamma)$ given in the sixth column of Table~\ref{tab:update-equations}. Next, we apply Theorem \ref{theorem:convergece} and \eqref{eq:update-sequence-limit} to update the estimate of $\hat{\gamma}_i$.
This ``fast'' update stage is the core of the \gls{fbsbl}.

\subsubsection*{Discussion of Theorem \ref{theorem:convergece}} 

Theorem~1 in \cite{shutin2011TSP:fastVSBL} analyzes the standard \gls{sbl} scenario, i.e., with $d=1$. Specifically, $2 d=2$ fixed points are computed and the locally stable one identified. In the \gls{bsbl} scenario, i.e., where $d>1$, several locally stable fixed points might exist since each of the polynomials $G_i(\gamma)$ given in the sixth column of Table~\ref{tab:update-equations} can have up to $P$ positive roots with $P$ denoting its degree.
Specifically, the polynomials $G_i(\gamma)$ given in rows ii and iii of this table have degree $P=2 d+1$, the polynomial in Row~iv has degree $P=2d$, and that in Row~v has degree $P=2d-1$.
The criterion \eqref{eq:update-sequence-limit} in our Theorem \ref{theorem:convergece} precisely specifies the fixed point toward which the sequence $\{\hat{\gamma}_i^{[n]}\}_{n=1}^{\infty}$ converges for any $d\geq 1$, i.e., including the classical case $d=1$.

\section{Equivalence Between BSBL Implementations}
\label{sec:equivalence}

Reference \cite{palmerNIPS2005} shows the equivalence between \gls{em}-based \gls{sbl} and \gls{vb}-based \gls{sbl}. 
In this section, we generalize this result
to \gls{bsbl}. Specifically, we show that the \gls{vb}-based implementation of \gls{bsbl} derived in Section~\ref{sec:variational-fast-solution} coincides with \gls{em}-based \gls{bsbl}, see e.g., \cite{zhang2011STSP:bSBL,zhang2013TSP:BlockSparseSBL}.
A direct consequence of this result, stated in Corollary~\ref{corollary:fast-bsbl-equivalence}, is that \gls{fbsbl} coincides with the coordinate-ascent-based implementation of \gls{bsbl}, see e.g., \cite{ma2017TSP:FastBSBL}.

First, we prove the next lemma. The proof is similar to that of  
\cite[Theorem~1]{palmerNIPS2005}. Throughout this section, we assume that the noise precision $\lambda$ is fixed and known.
\begin{lemma} \label{lemma:convex-representation}
The multivariate \gls{pdf} $p(\tilde{\bm{x}})=h \circ z(\tilde{\bm{x}})$, where $h(z)=\exp(-g(z^2))$ and $z(\tilde{\bm{x}})=\sqrt{\tilde{\bm{x}}^\hermitian\tilde{\bm{D}}\tilde{\bm{x}}}$ for some positive-definite matrix $\tilde{\bm{D}}$, can be represented in the convex variational form
\begin{align}\label{eq:prior-convex-type}
p(\tilde{\bm{x}}) = \sup_{\gamma > 0}\mathrm{N}\big(\tilde{\bm{x}};\,\bm{0},(\gamma\tilde{\bm{D}})^{-1}\big) \varphi(\gamma)
\end{align}
if, and only if, $g(z)$ is non-decreasing and concave on $(0,\infty)$. In this case, we can use the function
\begin{equation}
\varphi(\gamma) = \big|(\rho \gamma/\pi)\tilde{\bm{D}}\big|^{-\rho} \exp\big(g^{\ast}(\rho \gamma ) \big)
\end{equation}
where $g^{\ast}$ is the concave conjugate of $g$.
\end{lemma}

\begin{proof}
Applying \cite[Theorem~1]{palmerNIPS2005} to $h(z)$ yields
\begin{align}\label{eq:convex-condition-2}
h(z) = \sup_{\xi > 0} \mathrm{N}(z;\, 0,\, \xi^{-1})\sqrt{2\pi/\xi}\exp\big(g^{\ast}(\xi/2)\Big) \, .
\end{align} 
Inserting \eqref{eq:convex-condition-2} into the composition $p(\tilde{\bm{x}})=h \circ z(\tilde{\bm{x}})$ yields \eqref{eq:prior-convex-type} after substituting $\xi = 2\rho \gamma$ and some algebraic manipulations.
\end{proof}

Let $p(\bm{x})=\prod_{i=1}^{K}p(\bm{x}_i)$ where the \gls{pdf} $p(\tilde{\bm{x}})$ that occurs $K$ times in the right-hand factors admits the representation \eqref{eq:prior-convex-type}.
By omitting the supremum operator in \eqref{eq:prior-convex-type} for all $i=1,\ldots,K$
we obtain a lower bound on the evidence $p(\bm{y}) = \int p(\bm{y}|\bm{x})p(\bm{x})\,\mathrm{d}\bm{x}$ as follows:
\begin{align} \label{eq:em-lower-bound}
\tilde{p}(\bm{y};\bm{\gamma}) =
\int p(\bm{y}|\bm{x})\prod_{i=1}^{K} p(\bm{x}_i;\gamma_i)\varphi(\gamma_i)\,\mathrm{d}\bm{x} \leq p(\bm{y})
\end{align} 
where $p(\bm{x}_i;\gamma_i) = \mathrm{N}\big(\bm{x}_i;\,\bm{0},(\gamma_i\bm{D})^{-1}\big)$.
We seek the value of $\bm{\gamma}$ that, given $\bm{y}$, maximizes the lower bound $\tilde{p}(\bm{y};\bm{\gamma})$ and use the \gls{em}-algorithm to compute it.
For $\varphi(\gamma)=1$, $\tilde{p}(\bm{y};\bm{\gamma})$ coincides with the marginal likelihood that is maximized in \cite{ma2017TSP:FastBSBL} and \cite{zhang2013TSP:BlockSparseSBL} using coordinate ascent and the \gls{em} algorithm, respectively.

The \gls{em} algorithm returns a sequence of estimates of $\bm{\gamma}$ by successively maximizing the objective function
\begin{align} \label{eq:em-Q-function}
Q\big(\bm{\gamma},q_{\bm{x}}\big) = \Big<\ln \frac{p(\bm{y}|\bm{x})\prod_{i=1}^{K} p(\bm{x}_i;\gamma_i)\varphi(\gamma_i)}{q_{\bm{x}}(\bm{x})}\Big>_{q_{\bm{x}}}
\end{align}
with respect to $\bm{\gamma}$ and the proxy \gls{pdf} $q_{\bm{x}}$. 
The following theorem states the equivalence between the \gls{em}-based and \gls{vb}-based implementations of \gls{bsbl} in case $p(\gamma)$ and $\varphi(\gamma)$ are chosen such that they result in the same \gls{pdf} $p(\tilde{\bm{x}})$ according to \eqref{eq:prior-weights-type-I} and \eqref{eq:prior-convex-type}, respectively, for any $i=1,\ldots,K$.

\begin{theorem}\label{theorem:em-equivalent}
The proxy \gls{pdf} $q_{\bm{x}}^{\mathrm{EM}}$ maximizing \eqref{eq:em-Q-function} is equal to $q^\star_{\bm{x}}$ in \eqref{eq:weight-proxypdf} 
when $\bm{\gamma}$ in the former equation and $\hat{\bm{\gamma}}$ in the latter are set equal.
Similarly, the value $\bm{\gamma}^{\mathrm{EM}}$ maximizing \eqref{eq:em-Q-function} coincides with $\hat{\bm{\gamma}}$ computed from \eqref{eq:expectation-gamma-inverse-general} when in \eqref{eq:em-Q-function} and in \eqref{eq:precision-distribution} $q_{\bm{x}}$ and $q^\star_{\bm{x}}$, respectively, are equal.
Specifically, $\bm{\gamma}^{\mathrm{EM}}=[\gamma_1^{\mathrm{EM}} \ldots \gamma_K^{\mathrm{EM}}]^\transpose$ with
\begin{equation}\label{eq:theroem-em-vb-equivalence}
\gamma_i^{\mathrm{EM}} = \big<\gamma_i\big>_{q^\star_{\gamma_i}}=
	\hat{\gamma}_i = \omega_i^{\prime}(v_i)
\end{equation}
where 
\begin{equation}\label{eq:theroem-em-vb-equivalence-2}
\omega_i(v_i) = \inf_{\gamma_i>0} \big\{ v_i \gamma_i - \rho d \ln \gamma_i - \ln \varphi(\gamma_i) \big\}   
\end{equation}
with $v_i=\rho \big<\bm{x}_i^\hermitian\bm{D}\bm{x}_i\big>_{q_{\bm{x}}}$, $i=1,\ldots,K$ and $(\cdot)^\prime$ denotes the derivative.
\end{theorem}
\begin{proof}
The proof of the first part of the theorem is straightforward.
The proof of \eqref{eq:theroem-em-vb-equivalence}
is found in Appendix~\ref{sec:appendix:em-vb-equivalence}.
\end{proof}

\begin{corollary}\label{corollary:fast-bsbl-equivalence}  The stable fixed points of the fast \gls{vb} update
\begin{align}
\big\{\gamma^\ast \in \mathbb{R}_{++} : f_i(\gamma^\ast)-\gamma^\ast=0,\  \big|f_i^\prime(\gamma)\big|_{\gamma=\gamma^\ast} < 1\big\}
\end{align}
are the local maxima of the $i$th section $\gamma_i \mapsto \tilde{p}(\bm{y};\bm{\gamma})$ of $\tilde{p}(\bm{y};\bm{\gamma})$.
\end{corollary}

Note that a fixed point $\gamma^\ast$ of the recurrent relation $\hat{\gamma}_i^{[n+1]}=f_i(\hat{\gamma}_i^{[n]})$ is locally stable if $\big|f_i^\prime(\gamma)\big|_{\gamma=\gamma^\ast}<1$ \cite{shutin2011TSP:fastVSBL}.

\begin{proof}
The \gls{em} updates of $\bm{\gamma}$ are guaranteed to increase the value of $\tilde{p}(\bm{y};\bm{\gamma})$. Theorem~\ref{theorem:em-equivalent} states that the \gls{em} and \gls{vb} updates of $\bm{\gamma}$ are identical.
Hence, each \gls{vb} update of $\bm{\gamma}$ according to \eqref{eq:variational-update} increases the value of $\tilde{p}(\bm{y};\bm{\gamma})$ and so does the sequence $\{\hat{\gamma}_i^{[n]}\}_{n=1}^{\infty}$.
Consequently, a necessary condition for $\lim_{n\rightarrow\infty} \hat{\gamma}_i^{[n]} = \gamma_i$ is that $\gamma_i$ be a stationary point of $\tilde{p}(\bm{y};\bm{\gamma})$, with the local maxima coinciding with the locally stable fixed points.
\end{proof}
We show in Appendix~\ref{sec:appendix:conditions-convex-representation} that $p(\tilde{\bm{x}})$ in \eqref{eq:ghd} 
admits the convex-variational representation \eqref{eq:prior-convex-type}, for $(a,b,c)\in\Theta$, see \eqref{eq:Theta}.
Hence, Theorem~\ref{theorem:em-equivalent} and Corollary~\ref{corollary:fast-bsbl-equivalence} state that (i) the \gls{em}-based and \gls{vb}-based implementations of \gls{bsbl} are identical, and (ii) also \gls{bsbl} using coordinate ascent to maximize the marginal likelihood, e.g., \cite{ma2017TSP:FastBSBL}, and \gls{fbsbl} are identical as well.
As a result of (i), the presented fast solution can also be applied to \gls{em}-based \gls{bsbl} implementations, e.g., \cite{zhang2011STSP:bSBL,zhang2013TSP:BlockSparseSBL}. 
Note that inserting Jeffrey's density $p(\gamma)=\gamma^{-1}$ in the integral representation \eqref{eq:prior-weights-type-I} yields an improper density $p(\tilde{\bm{x}}) \propto \big(\frac{1}{\tilde{\bm{x}}^\hermitian\bm{D}
\tilde{\bm{x}}}\big)^{d/2}$ that also results from setting $\varphi(\gamma)=1$ in the convex representation \eqref{eq:prior-convex-type}.
Thus, the resulting \gls{em} and \gls{vb} updates must be identical according to Theorem~\ref{theorem:em-equivalent}. 
Indeed, the update rule \cite[Eq. (4)]{zhang2013TSP:BlockSparseSBL} derived from \gls{em} coincides with the update rule $\hat{\gamma}_i=d/\big<\bm{x}_i^\hermitian\bm{D}\bm{x}_i\big>_{q_{\bm{x}}^\star}$ of the \gls{vabsbl} algorithm using Jeffrey's density.
    \footnote{When comparing our work with \cite{zhang2013TSP:BlockSparseSBL}, note that we use precision hyperparameters $\gamma_i$ while \cite{zhang2013TSP:BlockSparseSBL} uses variances $\gamma_i^{-1}$ instead.}
Furthermore, an illustrative example of Corollary~\ref{corollary:fast-bsbl-equivalence} is given in Appendix~\ref{sec:appendix:equivalence-prove}, which explicitly derives that the \gls{bsbl} implementation using coordinate ascent to maximize the marginal likelihood obtained from \eqref{eq:em-lower-bound} with $\varphi(\gamma)=1$, e.g., \cite{ma2017TSP:FastBSBL,luessi2013TSP:VariationalBSBL}, and the presented \gls{fbsbl} algorithm using Jeffrey's density are identical.

\section{The F-BSBL Algorithm}
\label{sec:algorithm}

Note that rows and columns of $\hat{\bm{\Sigma}}$ in \eqref{eq:weights-mean} that correspond to a block $i$ with $\hat{\gamma}_i=\infty$ are zero and can be therefore discarded. 
As a result, $\hat{\bm{x}}_i$ will also be zero.
Algorithm~1 gives the pseudo-code of a computationally efficient algorithm akin to the original ``fast'' \gls{sbl} \cite{tipping2003WAIS:FastMarginalSparseBayesian} that combines the criterion for convergence/divergence of Theorem~\ref{theorem:convergece} with a ``bottom-up'' scheduling strategy that starts with an empty model, i.e., with all blocks deactivated by setting $\hat{\gamma}_i=\infty$ for all $i=1,\ldots,K$, resulting in an initial covariance matrix $\hat{\bm{\Sigma}}$ consisting of all zeros. The algorithm successively iterates through all blocks and performs updates \eqref{eq:update-sequence-limit}.
After cycling through all $i=1,\dots,K$, the algorithm updates the estimated mean of the noise precision using \eqref{eq:lambda-update-rule}.
These two steps are repeated until convergence or a maximum step number is reached.
	\footnote{The Matlab code for the \gls{fbsbl} algorithm is available at \url{https://doi.org/10.3217/t39cw-vrg36}.}
Since the algorithm starts with an empty model and parsimoniously adds new blocks of weights to the model, the matrices required in the computation typically retain small effective sizes (ignoring all rows/columns corresponding to zero weights), which significantly reduces computational complexity.

Special consideration must be given to the initialization of the algorithm when using the generalized inverse Gaussian prior with the setting $a=b=0$ and $c>0$, see Row~iv of Table~\ref{tab:update-equations}.
In this case, $f_i(\gamma)>\gamma$ for any large enough value of $\gamma$ (see \eqref{eq:theorem-sparsity-5}), which implies that any large enough initial estimate $\hat{\gamma}_i^{[0]}$ results in a sequence $\{\hat{\gamma}_i^{[n]}\}_{n=0}^{\infty}$ computed with the ``ad infinitum'' repeated update stage of Section~\ref{sec:variational-fast-solution} that diverges to infinity, regardless of the data $\bm{y}$ and dictionary $\bm{\Phi}$.
Hence, if the \gls{fbsbl} algorithm is initialized with $\hat{\gamma}_i=\infty$ for all $i=1,\cdots,K$, then \eqref{eq:update-sequence-limit} yields $\hat{\gamma}_i=\infty$, $i=1,\cdots,K$, and the algorithm returns the trivial result $\hat{\bm{x}}=\bm{0}$.
To avoid this, we propose to modify the update procedure of the hyperparameter estimates in the first three iterations of the algorithm based on the set of fixed points $\mathcal{G}_i$, $i=1,\ldots,K$ as follows. The estimate $\hat{\gamma}_i$ is set equal to the smallest fixed point of the update function $f_i$ regardless of whether $f_i(\hat{\gamma}_i)>\hat{\gamma}_i$ or not.
Specifically, $\hat{\gamma}_i=\min \mathcal{G}_i$ if $\mathcal{G}_i\neq \emptyset$, and $\hat{\gamma}_i=\infty$ otherwise. Note that these modified updates might decrease the \gls{elbo} in \eqref{eq:elbo}. After carrying out these first three iterations the algorithm proceeds with regular fast updates according to \eqref{eq:update-sequence-limit}, which ensures its convergence.

\begin{algorithm}[tbp]
\renewcommand{\algorithmicrequire}{\textbf{Input:}}
\renewcommand{\algorithmicensure}{\textbf{Output:}}
\caption{F-BSBL}
\label{alg:fastBSBL}
\begin{algorithmic}
\REQUIRE Observations $\bm{y}$, dictionary $\bm{\Phi}$, precision matrices $\bm{D}$.
\ENSURE Weights $\hat{\bm{x}}$, hyperparameters $\hat{\bm{\gamma}}$, noise precision $\hat{\lambda}$.

\STATE Initialize $n=1$, $\hat{\lambda} = \frac{2N}{\|\bm{y}\|^2}$ and $\hat{\gamma}_i = \infty$ for $i=1,\dots,K$.%

\WHILE{not converged}
\FORALL{$i = 1,\,\dots,\, K$}
\STATE $\hat{\bm{\Sigma}}_{\sim i}\leftarrow(\hat{\lambda}\bm{\Phi}^\hermitian\bm{\Phi} + \sum_{k=1,k\neq i}^{K} \hat{\gamma}_k\bm{E}_k\bm{D}\bm{E}_k^\transpose)^{-1}$.
\STATE Calculate $\bm{U}_i$, $\bm{S}_{i}$ and $\bm{q}_{i}$ (See Appendix~\ref{sec:variational-fast-solution:update-polynomials}).%
\STATE $\mathcal{G}_i \leftarrow $ Set of positive roots of $G_i(\gamma)$.

\IF {$n \leq 3$}
\STATE $\hat{\gamma}_i\leftarrow \min \mathcal{G}_i$ if $\mathcal{G}_i\neq \emptyset$, else $\infty$.
\ELSE
\STATE Update $\hat{\gamma}_i$ using \eqref{eq:update-sequence-limit}.
\ENDIF

\ENDFOR
\STATE $\hat{\bm{\Sigma}}\leftarrow(\hat{\lambda}\bm{\Phi}^\hermitian\bm{\Phi} + \hat{\bm{\Gamma}})^{-1}$.
\STATE $\hat{\bm{x}} \leftarrow \hat{\lambda}\hat{\bm{\Sigma}}\bm{\Phi}^\hermitian\bm{y}$.
\STATE Update $\hat{\lambda}$ using \eqref{eq:lambda-update-rule}.
\STATE $n \leftarrow n+1$.
\ENDWHILE		
\end{algorithmic}
\end{algorithm}

\subsubsection*{Computational Complexity}
Let $\|\cdot\|_0$ denote the $\ell_0$-quasi-norm, such that $\hat{M}=\|\hat{\bm{x}}\|_0$ is the (estimated) number of nonzero weights. Assuming that $d$ is a small constant compared to $\hat{M}$, the computational complexity of the relevant steps in the algorithm is as follows:
\begin{itemize}
\item Calculating the coefficients of the polynomials $A_i(\gamma)$ and $B_i(\gamma)$ has complexity $\mathcal{O}(\hat{M}^3)$.
\item Solving for the roots of the polynomial $G_i(\gamma)$, e.g., via the eigenvalues of the companion matrix \cite{edelman1995polynomial}, has complexity $\mathcal{O}(d^3)$.
\item Updating $\hat{\gamma}_i$ according to Theorem \ref{theorem:convergece} with $\mathcal{G}_i$ already obtained has complexity $\mathcal{O}(d)$.
\end{itemize}
The computationally most complex operation is the computation of the coefficients of $A_i(\gamma)$ and $B_i(\gamma)$, since it requires the calculation of the matrix $\hat{\bm{\Sigma}}_{\sim i}$ defined in Appendix~\ref{sec:variational-fast-solution:update-polynomials}. Only rows and columns of $\hat{\bm{\Sigma}}_{\sim i}$ corresponding to nonzero blocks ($\hat{\gamma}_j<\infty$, $j\neq i$) must be considered. Hence, this operation has complexity $\mathcal{O}(\hat{M}^3)$.
Due to its inherent structure, the \gls{em}-based algorithm proposed in \cite{zhang2013TSP:BlockSparseSBL} computes finite estimates of the hyperparameters.
In its practical implementation, the algorithm is augmented by including a pruning procedure that deactivates components with a hyperparameter estimate larger than a specified threshold. However, it typically takes several iterations of the algorithm until some hyperparameter estimates increase beyond this threshold, and therefore the corresponding components are deactivated. Thus, the matrix inversion operation carried out in this augmented scheme has complexity close to $\mathcal{O}(N^3)$ during the first iterations.
It follows that for $\hat{M} < N$ (i.e., if the estimate is sparse) the \gls{fbsbl} algorithm has lower computational complexity than the \gls{em}-based algorithm in \cite{zhang2013TSP:BlockSparseSBL}.
Furthermore, the fast update procedure in Theorem~\ref{theorem:convergece} is the culmination of (infinitely) many individual updates. 
Hence, the \gls{fbsbl} algorithm typically requires fewer iterations to achieve convergence than the \gls{em}-based algorithm, see Section \ref{sec:results}.

\section{Analysis of F-BSBL}
\label{sec:analysis}

\subsection{How to Select the Prior Parameters $a$, $b$ and $c$?}
\label{sec:analysis:sparsity}

Divergence of $\hat{\gamma}_i$ for many $i=1,\dots,K$ is key to both obtaining a sparse result and the computational advantage of the \gls{fbsbl} algorithm.
To obtain insights into the conditions that lead to convergence or divergence of the sequence $\{\hat{\gamma}_i^{[n]}\}_{n=0}^{\infty}$ generated by $\hat{\gamma}_i^{[n+1]}=f_i(\hat{\gamma}_i^{[n]})$, we analyze the update function $f_i(\gamma)$ as $\gamma \rightarrow \infty$.

\begin{lemma}
\label{lemma:asymptotic-relation-Bi-Ai}
The rational function defined in Lemma~\ref{lemma:expectation-as-polynomial-fraction} behaves according to
$B_i(\gamma)/A_i(\gamma) = d/\gamma + o(\gamma^{-2})$ as $\gamma\rightarrow \infty$.
\end{lemma}
Here, $o(\cdot)$ is the little-o notation.
\begin{proof}
From \eqref{eq:hi-sum} in Appendix~\ref{sec:appendix:derivative-hi} we obtain
\begin{align}
	\frac{B_i(\gamma)}{A_i(\gamma)} &= \gamma^{-1}\big[ d + o(1)\gamma^{-1} \big] & \text{as }& \gamma\rightarrow \infty
\end{align}
after some algebraic manipulations.
\end{proof}

\begin{theorem}
\label{theorem:sparsity}

For any $b\in \mathbb{R}_{+}$, provided that either $a > 0$ or $c < 0$ the sequence $\{\hat{\gamma}_i^{[n]}\}_{n=1}^{\infty}$ always converges regardless of the measurement vector $\bm{y}$ or dictionary $\bm{\Phi}$.
\end{theorem}

Theorem~\ref{theorem:sparsity} is equivalent to the statement that both $a=0$ and $c\geq 0$ are necessary conditions such that the sequence $\{\hat{\gamma}_i^{[n]}\}_{n=1}^{\infty}$ 
may diverge depending on $\bm{y}$ and $\bm{\Phi}$.
In other words, by selecting the prior density \eqref{eq:generalized-inverse-gaussian} with $a=0$ and $c\geq 0$, the resulting estimator embodies an inherent threshold of the estimated weights that in effect switches off sufficiently weak components, whereas the selection $a>0$ or $c<0$ leads to an estimator that, although it induces a bias towards zero for small weight estimates, never eventually set them to zero.

\begin{proof}
A sufficient condition for $\{\hat{\gamma}_i^{[n]}\}_{n=1}^{\infty}$ to converge is that any sufficiently large initial value $\hat{\gamma}_i^{[0]}$ results in a decreasing sequence, i.e., that 
\begin{align}
	\label{eq:sufficient-condition-convergence}
	f_i(\gamma) - \gamma < 0
\end{align}
holds for any $\gamma$ sufficiently large.

\subsubsection*{Case I, $b > 0$ and $a > 0$} Using Lemma~\ref{lemma:expectation-as-polynomial-fraction} and \eqref{eq:gig-param-ai-update}, we write \eqref{eq:update-function-general-form} as
\begin{align}
	\label{eq:theorem-sparsity-1}
	f_i(\gamma) = b\, u_i(\gamma)^{-1/2} \,\frac{K_{\hat{c}+1}(\sqrt{u_i(\gamma)})}{K_{\hat{c}}(\sqrt{u_i(\gamma)})}
\end{align}
where $u_i(\gamma) = b\big(a + 2\rho B_i(\gamma)/A_i(\gamma)\big)$.
Applying Lemma~\ref{lemma:asymptotic-relation-Bi-Ai}, we find that $u_i(\gamma)$ converges to $ab$ as $\gamma \rightarrow \infty$. Hence, $f_i(\gamma)$ approaches some finite value as $\gamma\rightarrow \infty$, resulting in 
\begin{align}
	\lim_{\gamma \rightarrow\infty} f_i(\gamma) - \gamma = -\infty 
	\, .
\end{align}

\subsubsection*{Case II, $b>0$ and $a=0$}
In this case, $u_i(\gamma)$ converges to $0$ as $\gamma \rightarrow \infty$ by Lemma~\ref{lemma:asymptotic-relation-Bi-Ai}. 
From \cite[Eq. (9.6.9)]{abramowitz1972:handbookMathFun}, we have the asymptotic equivalence
\begin{align}
	\label{eq:theorem-sparsity-3}
	\frac{K_{\hat{c}+1}(z)}{K_{\hat{c}}(z)} &\sim \frac{2\hat{c}}{z} & \text{as } z \rightarrow 0
	\, .
\end{align}
We set $z=\sqrt{u_i(\gamma)}$ in \eqref{eq:theorem-sparsity-3}, insert the latter into \eqref{eq:theorem-sparsity-1}, and apply the definitions of $A_i(\gamma)$ and $B_i(\gamma)$ given in \eqref{eq:polynom-A} and \eqref{eq:polynom-B}, respectively, to find
\begin{align} \label{eq:theorem-sparsity-4}
	f_i(\gamma) - \gamma &= \frac{c}{\rho d}\gamma + o(1) & &\text{as } \gamma \rightarrow \infty
	\, 
\end{align}
which fulfills the sufficient condition \eqref{eq:sufficient-condition-convergence} if $c < 0$.

\subsubsection*{Case III, $b=0$ and $a>0$}
Using the simplified expression for $f_i$ given in the third row of Table~\ref{tab:update-equations}, we find that
\begin{align}
	\label{eq:theorem-sparsity-2}
	f_i(\gamma) - \gamma
	&= -\gamma + o(1) & & \text{as } \gamma\rightarrow \infty
\end{align}
after a few algebraic manipulations.

\subsubsection*{Case IV, $b=0$ and $a=0$} Using the definition of $f_i$ given in the fourth row of Table~\ref{tab:update-equations}, we find
\begin{align} \label{eq:theorem-sparsity-5}
	f_i(\gamma) - \gamma 
	&=\frac{c}{\rho d}\gamma + o(1) & &\text{as } \gamma \rightarrow \infty
\end{align}
same as for the case $b > 0$.

In summary, the sufficient condition \eqref{eq:sufficient-condition-convergence} is always fulfilled in Case I and III (i.e., when $a>0$), whereas it is fulfilled if, and only if, $c<0$ for Case II and IV.
\end{proof}

Theorem~\ref{theorem:sparsity} states that priors with $a=0$ or $c<0$ do not introduce any pruning and, thus, do not yield a sparse estimate of the weight vector. For $b>0$ the set of fixed points $\mathcal{G}_i$ can only be obtained in closed form if $c=-\rho d -\frac{1}{2}$. Thus, we focus primarily on the case $a=b=0$ in the remainder of this work. This setting yields the improper density $p(\gamma)= \gamma^{c-1}$, $c \in [0, \infty)$ corresponding to rows iv and v of Table~\ref{tab:update-equations} for $c>0$ and $c=0$, respectively.

\subsection{Modified Threshold}
\label{sec:analysis:threshold}

When \gls{sbl} is applied to line spectrum estimation to detect multipath components and estimate their dispersion parameters, e.g. their time and direction of arrival, in radio communication, it is known to overestimate the number of components \cite{badiu2017TSP:VSBL, hansen2018TSP:SuperFastLSE, leitinger2020Asilomar, GreLeiWitFle:TWC2024}.
A common solution to this problem is to modify the threshold in the pruning procedure of \gls{fbsbl}, as done in \cite{shutin2011TSP:fastVSBL}.

It can be readily shown that the sequence $\{\hat{\gamma}_i^{[n]}\}_{n=1}^{\infty}$ generated by ``ad infinitum'' execution of the update stage described in Section~\ref{sec:variational-fast-solution}
converges in practice to a \emph{locally stable} fixed point $\gamma^\ast \in \mathcal{G}_i$, i.e., fulfilling
\begin{align} \label{eq:fixed-point-condition-locally-stable}
\big|f_i^\prime(\gamma)\big|_{\gamma=\gamma^\ast} < 1
\, .
\end{align}
Reference \cite{shutin2011TSP:fastVSBL} shows that for $d=1$ and using Jeffrey's density ($a\rmv\rmv=\rmv\rmv b\rmv\rmv = \rmv\rmv c \rmv\rmv =\rmv\rmv 0)$ as hyperprior, condition \eqref{eq:fixed-point-condition-locally-stable} is equivalent to $|q_{i,1}|^2/s_{i,1} \geq 1$.
The authors suggest to introduce a heuristic threshold $|q_{i,1}|^2/s_{i,1} > \tilde{\chi}$ with $\tilde{\chi}\geq 1$ instead.
They verify the effectiveness of this method through numerical simulations.
For $d>1$, the relation between condition \eqref{eq:fixed-point-condition-locally-stable} and the values of $q_{i,l}$ and $s_{i,l}$, $l = 1,\,\dots,\,d$ is more involved.
Nevertheless, we introduce a threshold $\chi \leq 1$ and retain among the fixed points in $\mathcal{G}_i$ only those 
fulfilling
\begin{equation} \label{eq:threshold}
\big|f_i^\prime(\gamma)\big|_{\gamma=\gamma^\ast} < \chi
\, .
\end{equation}
A disadvantage of this approach is that for $\chi < 1$ the fast update stage might not produce a hyperparameter value identical to the limit of a sequence generated with the ``ad infinitum'' executed update stage in Section~\ref{sec:variational-fast-solution}.
Hence, any guarantee that the \gls{elbo} is increased in each step and that the algorithm converges is lost.
The unmodified update rule \eqref{eq:update-sequence-limit} is recovered by setting $\chi=1$, in which case the algorithm is guaranteed to increase the \gls{elbo} in each iteration and, thus, to converge.

As discussed in Subsection~\ref{sec:analysis}, for the setting $a \rmv\rmv = \rmv\rmv b \rmv\rmv = \rmv\rmv 0$ and $c>0$ (see Row~iv of Table~\ref{tab:update-equations}), increasing $c$ results in the hyperprior placing more probability mass on its tail. Since a hyperparameter scales the prior precision matrix of its associated block, this can be interpreted as an additional bias that drives the estimates of the weights of this block towards zero and, thus, increases the interval in which these estimates are set to zero.
Hence, both $c$ and $\chi$ can be used to tune the threshold in the inherent pruning procedure of the \gls{fbsbl} algorithm.

\subsection{Simulation Study}
\label{sec:analysis:simulation}

In order to verify the theoretical analysis of Section \ref{sec:analysis:sparsity} and to investigate the validity of the approach of Section \ref{sec:analysis:threshold}, we conduct an experiment that illustrates the pruning behavior as a function of the selected hyperprior parameters ($a$, $b$, $c$).
A noiseless system is considered with an identity measurement matrix, i.e., $M=N$ and $\bm{\Phi}=\bm{I}_N$, consisting of a single block, i.e., $K=1$ and $d=N$, which results in the trivial model $\bm{y}=\bm{x}$.
Since $\bm{\Phi}=\bm{I}_{N}$, a perfect estimator for the noiseless case would be $\hat{\bm{x}}=\bm{y}$. However, to obtain a sparse estimate in more practical scenarios, ``weak'' components should be pruned, i.e., estimates $\hat{\bm{x}}$ with norm $\|\hat{\bm{x}}\|$ small compared to the assumed/estimated noise variance $1/\hat{\lambda}$ should be set to zero by the estimator.

\begin{figure}
\centering
\setlength{\figurewidth}{0.4\textwidth}
\setlength{\figureheight}{2.5cm}
\def\plotlinewidth{1.5pt}
\def\datapath{./pgf}
\includegraphics{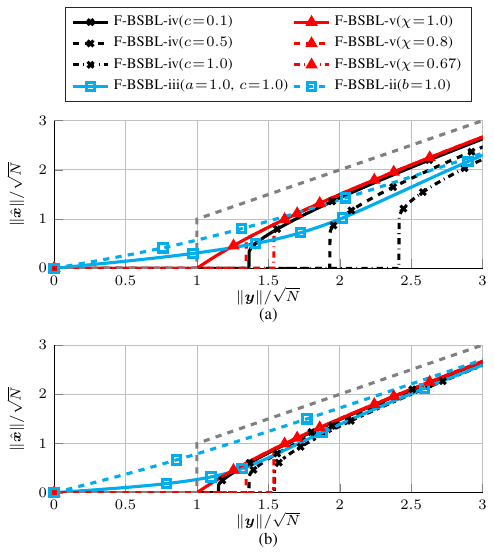}
\caption{Thresholding function of the \gls{fbsbl} algorithm for different settings of the parameters of the generalized inverse Gaussian prior under the assumption of an identity measurement matrix, i.e., $\bm{\Phi}=\bm{I}_N$, and a single block, i.e., $K=1$ and $d=N$: (a) $d=N=2$, (b) $d=N=10$.}
\label{fig:thresholding}
\end{figure}

Let $\bm{1}_{N}$ denote a vector of ones with length $N$. Weights $\bm{x}=\alpha \bm{1}_{N}$ and their corresponding (noiseless) measurements $\bm{y}=\bm{x}$ are generated repeatedly using different values of the scale $\alpha$.
We evaluate the fast update rule for $\hat{\gamma}_1$ and the resulting weight estimate $\hat{\bm{x}}=\bm{y}/(1+\hat{\gamma}_1)$ depending on $\alpha$, assuming $\hat{\lambda}=1$.
Figure \ref{fig:thresholding}a and \ref{fig:thresholding}b depict the normalized norm $\|\hat{\bm{x}}\| / \sqrt{N}$ of the weights $\hat{\bm{x}}$ versus the normalized norm of $\bm{y}$ to illustrate the thresholding functions. ``Hard'' thresholding, i.e., $\hat{\bm{x}} = \bm{y}$ if $\|\bm{y}\|/\sqrt{N}>1$ and $\hat{\bm{x}}=\bm{0}$ otherwise, is represented by the dashed gray lines in Figure~\ref{fig:thresholding}. We use lowercase roman numerals to denote the respective row of Table~\ref{tab:update-equations} used for the prior, e.g., \gls{fbsbl}-iii($a=1,b=1$) refers to the \gls{fbsbl} algorithm with setting $b=0$, yielding the hyperprior density $p(\gamma)=\gamma^{c-1}\exp\{-\frac{a}{2}\gamma\}$ given in Row~iii of Table~\ref{tab:update-equations}.
Unless otherwise stated, the threshold $\chi=1$ is used.

\Gls{fbsbl}-ii and \gls{fbsbl}-iii do not inherently include a pruning condition, since $\hat{\bm{x}}=\bm{0}$ if and only if $\bm{y}=\bm{0}$. This is in line with both Theorem~\ref{theorem:sparsity} and similar results from the literature \cite{babacan2014TSP:variationalBSBL, pedersen2015SP}.
Comparing the estimators obtained from setting $a=b=0$ and $c\geq 0$ yielding densities of the form $p(\gamma) = \gamma^{c-1}$ corresponding to \gls{fbsbl}-iv for $c>0$ and \gls{fbsbl}-v for $c=0$, we observe that, indeed, these estimators show a pruning behavior.
That is, small (nonzero) values of $\bm{y}$ lead to $\hat{\bm{x}}=\bm{0}$ (i.e., $\hat{\gamma}_1=\infty$).
For \gls{fbsbl}-v, decreasing $\chi$ naturally increases the region where $\hat{\bm{x}}=\bm{0}$, i.e., the region in which the component is pruned.
A similar behavior is observed for \gls{fbsbl}-iv when increasing the value of $c$.
Note that increasing $c$ increases the bias towards zero.
Furthermore, it is noteworthy to highlight the effect of varying the block size on the solution ($d=2$ and $d=10$ shown in Figure \ref{fig:thresholding}a and \ref{fig:thresholding}b, respectively). For \gls{fbsbl}-v, the cutoff value remains constant. However, the likelihood of noise randomly interfering constructively or destructively with the estimated weights decreases as the block size increases, due to the decreasing likelihood of all weights within the block being affected in the same manner.
Consequently, a fixed threshold $\chi$ would result in the rate of erroneous block classifications varying with the block size.
In contrast, for \gls{fbsbl}-iv we observe a reduction of the cutoff region as the block size increases, compensating this effect to some extent.

Finally, from Figure \ref{fig:thresholding} we can see that both \gls{fbsbl}-iv and \gls{fbsbl}-v perform a similar pruning, and therefore achieve a similar estimation performance, when tuned equivalently, e.g., a block size of $d=10$ using either \gls{fbsbl}-iv($c=1$) or \gls{fbsbl}-v($\chi=0.67$).
We conclude that the use of a hyperprior density $p(\gamma)=\gamma^{c-1}$ with $c>0$ (\gls{fbsbl}-iv) is to be preferred over that of a hyperprior density with $c=0$ and including an additional threshold $\chi$ (\gls{fbsbl}-v) to retain the convergence guarantees of the \gls{vb} algorithm.

\section{Numerical Evaluation}
\label{sec:results}

To investigate the performance of the algorithm, we consider a real-valued scenario (i.e., $\rho=1/2$). We generate a dictionary matrix with $M=2N$ columns and assume that $N=200$ measurements are obtained unless otherwise stated. The elements of the dictionary matrix are drawn independently from a standard normal and each column of $\bm{\Phi}$ is normalized such that it has unit $\ell_2$ norm.
The corresponding weight vector $\bm{x}$ is partitioned into $M/d=40$ blocks of size $d=10$ each.
Among them a certain number of active blocks are randomly selected, such that the desired sparsity ratio $\delta=\frac{\|\bm{x}\|_0}{N}=0.2$ is achieved.
The weight vectors of active blocks are independently drawn from a multivariate Gaussian distribution with zero mean and identity covariance matrix, while they are set to zero for the non-active blocks.
The indices of the active blocks are unknown to the algorithm.
The noise precision $\lambda$ is chosen, such that the \gls{snr} defined as $\text{SNR}=\frac{\lambda\|\bm{\Phi}\bm{x}\|^2}{N}$ equals $15\,\text{dB}$.
As performance metric we use the \acrfull{nmse} defined as $\text{NMSE}=\frac{\|\bm{x}-\hat{\bm{x}}\|^2}{\|\bm{x}\|^2}$, averaged over 100 simulation runs.

For the \gls{fbsbl} algorithm, we use Jeffrey's prior for the noise precision, 
which is obtained by setting $\epsilon=\eta=0$ in \eqref{eq:noise-prior}.
For the hyperprior we set $a=b=0$. To illustrate the difference of using either $c$ or $\chi$ to tune the threshold used for the pruning of ``weak'' components, we consider two variants of \gls{fbsbl}: \gls{fbsbl}-iv($c=1$) uses the density in Row~iv of Table~\ref{tab:update-equations} with $c=1$ and the (\gls{sbl}) threshold $\chi=1$, while \gls{fbsbl}-v($\chi=0.67$) uses Jeffrey's density in Row~v of Table~\ref{tab:update-equations} (i.e., $c=0$) with $\chi=0.67$.

As benchmarks we use two \gls{em}-based implementations of \gls{bsbl} proposed in \cite{zhang2013TSP:BlockSparseSBL}, referred to as BSBL-EM and BSBL-BO respectively.
    \footnote{The code for the BSBL-EM and BSBL-BO algorithms was obtained from \url{http://dsp.ucsd.edu/~zhilin/BSBL.html}}
Note that the BSBL-EM algorithm is identical to the \gls{vabsbl} algorithm using Jeffrey's hyperprior.
For reference, we also include an oracle \gls{mmse} estimator, which is given the index of the active blocks and calculates the weights of the active blocks using the pseudoinverse of the corresponding columns of the dictionary.
The same convergence criterion is used for all algorithms. Let $\hat{\bm{\sigma}}=[\hat{\gamma}_1^{-1} \ist\iist \hat{\gamma}_2^{-1} \ist \cdots \ist \hat{\gamma}_K^{-1}]^\transpose$ be the vector of estimated weight variances. Each algorithm is considered to have converged if the number and indices of the estimated active blocks do not change from one iteration to the next and $\|\hat{\bm{\sigma}}^{[n]}-\hat{\bm{\sigma}}^{[n-1]}\|_1 / \|\hat{\bm{\sigma}}^{[n]}\|_1 < 10^{-4}$, where $\hat{\bm{\sigma}}^{[n]}$ is the value of $\hat{\bm{\sigma}}$ estimated at the $n$th iteration, and $\|\cdot\|_1$ denotes the $\ell_1$-norm.

\begin{figure*}
\centering
\setlength{\figurewidth}{0.25\textwidth}
\setlength{\figureheight}{2.3cm}
\def\datapath{./pgf/}
\includegraphics{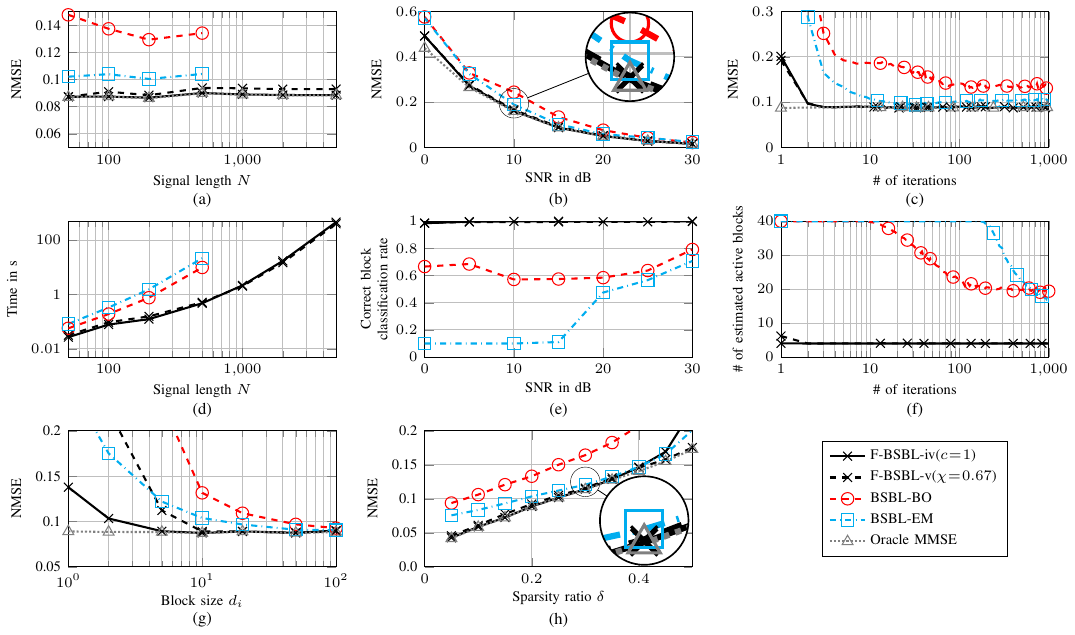}
\vspace*{-4mm}
\caption{Performance of the \gls{fbsbl} algorithm under the assumption of a dictionary matrix with standard Gaussian distributed entries. Results are averaged over 100 realizations. Depicted is the NMSE versus: the signal length (a), the SNR (b) the number of iterations (c), the block size (g), and the sparsity ratio (h).
Also reported are the runtime versus the signal length (d), the rate of correct block classifications versus the SNR (e), and the number of estimated active blocks versus the number of iterations (f).
Unless otherwise stated, the following parameters were used in the simulations: $N=200$, $M=2N$, $\text{SNR}=15\,\text{dB}$, $\delta=0.2$ and $d=10$.}
\label{fig:mainEval}
\end{figure*}

Figures \ref{fig:mainEval}a-\ref{fig:mainEval}c depict the \gls{nmse} of the algorithms versus the signal length $N$, the \gls{snr} and the number of iterations. We define the number of iterations as the number of times the main loop of the algorithm is executed, i.e., all estimates $\hat{\bm{x}}$, $\hat{\bm{\Sigma}}$, $\hat{\lambda}$ and $\hat{\gamma}_i$, $i=1,\dots,K$ are updated.
Figure \ref{fig:mainEval}d shows the runtime of the algorithms again versus the signal length $N$.
For $N=500$, the runtime of the \gls{fbsbl} algorithm is approximately 2 orders of magnitude smaller than that of the BSBL-EM and BSBL-BO algorithms \cite{zhang2013TSP:BlockSparseSBL}, while achieving an \gls{nmse} virtually identical to that of the oracle \gls{mmse} estimator. As shown in Figures \ref{fig:mainEval}b and \ref{fig:mainEval}c, the \gls{fbsbl} algorithm achieves an \gls{nmse} almost identical to that of the oracle estimator already after 3 iterations and over a wide range of \glspl{snr}.
Figure \ref{fig:mainEval}e shows the relative frequency of correctly classifying the hyperparameter estimates as converged or diverged (i.e., the relative frequency of classifying each block correctly as active or non-active). While the \gls{fbsbl} algorithm correctly classifies the blocks almost all the time, the BSBL-EM and BSBL-BO algorithms overestimate the number of active blocks (see Figure~\ref{fig:mainEval}f).
The erroneous classification of non-active blocks as active also leads to an increased \gls{nmse} of the BSBL-EM and BSBL-BO algorithms compared to the \gls{fbsbl} algorithm, as seen e.g., in Figure~\ref{fig:mainEval}a.

The difference in runtime of the algorithms can be explained by Figures \ref{fig:mainEval}c and \ref{fig:mainEval}f.
While the \gls{fbsbl} algorithm achieves an \gls{nmse} virtually identical to that of the oracle estimator after three iterations already, the BSBL-BO and BSBL-EM algorithms require far more iterations to converge.
Furthermore, as detailed in Figure \ref{fig:mainEval}f, the \gls{fbsbl} algorithm estimates the number of active blocks with almost perfect accuracy from the first iteration on.
Thus, the \gls{fsbl} algorithm requires few iterations, which themselves have low computational complexity due to the low effective dimension of 
$\hat{\bm{\Sigma}}$. Remember that entries in this matrix 
corresponding to deactivated blocks are exactly zero and can therefore be effectively discarded.
In contrast, it can be seen in Figure \ref{fig:mainEval}f, that the BSBL-BO and BSBL-EM algorithms keep all 40 blocks active for the first, respectively, 10 and 100 iterations before they start to deactivate them.
Thus, the BSBL-EM and BSBL-BO algorithms require a higher number of computationally more demanding iterations than the \gls{fsbl} algorithm to converge.

Additionally, we evaluate how varying the block size $d$ and the sparsity ratio $\delta$ affects the algorithm's performance in terms of \gls{nmse}.
We use $N=500$ and $N=200$ to evaluate the performance of the algorithm versus, respectively, the block size and the sparsity ratio.
Figures \ref{fig:mainEval}g and \ref{fig:mainEval}h show the \gls{nmse} resulting from these numerical experiments.
Here again, the \gls{fbsbl} algorithm outperforms the BSBL-EM and BSBL-BO algorithms for most considered sparsity ratios and \glspl{snr}. The runtime and correct block classification rate was evaluated as well.
However, these results are similar to those presented in Figures \ref{fig:mainEval} d and \ref{fig:mainEval} e.
Thus, they are omitted for the sake of brevity.
The results of the experiment with varying block size show that \gls{fbsbl}-iv is again superior to BSBL-EM and BSBL-BO, as well as to \gls{fbsbl}-v. The difference in performance between the two variants of \gls{fbsbl} stems from their respective erroneous block classification rate, specifically the rate of erroneously classifying non-active blocks as active, which is smaller for \gls{fbsbl}-iv, see Section \ref{sec:analysis:simulation}.

\section{Application Example: DOA Estimation}
\label{sec:application-doa}	
\subsection{System Model}

Consider a uniform linear array consisting of $N$ 
antenna elements which receive bandpass signals from $K$ sources located in its far field. Here, we use the complex baseband representation of signals, so the model is complex. The sources' signals originate from distinct \glspl{doa} that are assumed to belong to a grid $\{\theta_1,\ldots,\theta_k\}\subseteq[-90^\circ,90^\circ)$. By convention, $\theta_i$ is the \gls{doa} of the $i$th source, $i=1,\ldots,K$. 
Multiple measurements $\bm{y}_t\in\mathbb{C}^N$, $t=1,\dots,d$ are obtained, where $t$ denotes a time index.
These are modeled as
\begin{equation} \label{eq:doa:signal-model-single}
\bm{y}_t = \sum_{i=1}^{K} \bm{\psi}(\theta_i)s_i[t] + \bm{v}_t\quad ,\quad t=1,\dots,d
\,.
\end{equation}
In \eqref{eq:doa:signal-model-single}, $s_i[t]$ is the signal of the $i$th source, $i=1,\dots,K$ and
\begin{align}
\bm{\psi}(\theta)=\frac{1}{\sqrt{N}}\big[1\ e^{-j2\pi\frac{p_2}{\mu}\sin\theta}\ \cdots\  e^{-j2\pi\frac{p_N}{\mu}\sin\theta}\big]^\transpose
\end{align}
with $\mu$ denoting the wavelength of the sources' signals and $p_n$ denoting the distance from sensor 1 to sensor $n$, $n=1,\ldots,N$ is the steering vector of the uniform linear array in direction $\theta\in[-90^\circ,90^\circ)$. Finally, $\bm{v}_{t}$ is a vector-valued stationary, spatially and temporally white, Gaussian noise process. Thus, 
all entries of $\bm{v}_{t}$ for $t=1,\ldots,d$ are independent, identically distributed, zero-mean Gaussian random variables with variance $\lambda^{-1}$. 

An unknown number $L<K$ of sources are active and we aim to estimate $L$ along with the \glspl{doa} of these sources.
Let $\bm{x}_i=\big[s_i[1] \ist\cdots\ist s_i[d]\big]^\transpose$, $i=1,\dots,K$.
If $i$ is the index of an active source, $\bm{x}_i$ is zero-mean Gaussian with covariance $\bm{\Sigma}_{\text{s}}$.
If $i$ is the index of a non-active source then $\bm{x}_i=\bm{0}$.
Vectors of signals of distinct active sources are uncorrelated and follow the same distribution.
Defining $\bm{Y}=[\bm{y}_1\ \cdots \bm{y}_d]$, $\bm{X}=[\bm{x}_1\ \cdots\ \bm{x}_{K}]^\transpose$ and $\bm{\Psi}=[\bm{\psi}(\theta_1)\ \cdots\ \bm{\psi}(\theta_K)]$, we arrive at 
the signal model for multiple measurement scenarios
\begin{align} \label{eq:doa:mmv-model}
\bm{Y} = \bm{\Psi}\bm{X}+ \bm{V}.
\end{align}
We rearrange \eqref{eq:doa:mmv-model} into a block-sparse model (see Section~\ref{sec:signal-model:setup}) such that the \gls{fbsbl} algorithm can be applied to compute an estimate $\hat{\bm{X}}$ of $\bm{X}$.
We use the number of nonzero rows of $\hat{\bm{X}}$ as the estimate $\hat{L}$ of $L$ and obtain as \gls{doa} estimates the points of the grid used to compute the columns of $\bm{\Psi}$ corresponding to the nonzero rows of $\hat{\bm{X}}$.

\subsection{DOA Estimation Results}

Since the BSBL-EM and BSBL-BO algorithms from \cite{zhang2013TSP:BlockSparseSBL} are not applicable to a complex valued signal model, we compare the \gls{fbsbl} algorithm against the DOA-SBL algorithm proposed for multiple measurements in \cite{gerstoftSPL2016:SBLforDoA}.
\footnote{The code for the DOA-SBL algorithm was obtained from \url{https://github.com/gerstoft/SBL}.}
In this study, the actual weights $\bm{X}$ are of secondary interest. Thus, we resort to the \gls{ospa} metric \cite{schuhmacher2008TSP:OSPA} of the \glspl{doa} estimates to evaluate the performance of the algorithms.
The \gls{ospa} is a metric that considers both the actual estimation error as well as the number of erroneously classified blocks.
The parameters for the \gls{ospa} are the order and the cutoff-distance, which are set to $1$ and $5^{\circ}$, respectively.
Since the DOA-SBL algorithm does not directly estimate a sparse weight vector, we evaluate two versions of it.
The DOA-SBL algorithm estimates variance hyperparameters for each block of weights and the noise variance, which we denote as $\zeta_{i,\text{DOA-SBL}}$, $i=1,\ldots,K$ and $\sigma^2_{\text{DOA-SBL}}$, respectively.
    \footnote{The variance hyperparameters are denoted $\gamma_i$, $i=1,\dots,K$ in \cite{gerstoftSPL2016:SBLforDoA}.}
We define the estimated component \gls{snr} of the $i$th block as $\text{SNR}_{i,\text{DOA-SBL}} = \frac{\zeta_{i,\text{DOA-SBL}}}{\sigma^2_{\text{DOA-SBL}}}$.
The first variant of the algorithm returns as \gls{doa} estimates all grid points corresponding to blocks with estimated component \gls{snr} larger than the threshold.
	\footnote{The authors of \cite{gerstoftSPL2016:SBLforDoA} use a similar pruning procedure in a related paper \cite{park2023SValse}.}
Based on preliminary simulations, we adapted the threshold to maximize the performance of this variant in the actual scenarios. 
The second variant is an oracle version of the DOA-SBL algorithm which is given the true number of components $L$. 
It returns as \gls{doa} estimates the grid points 
corresponding to the $L$ blocks with largest estimated component \gls{snr}.
To account for the increasing processing gain at increasing arrays sizes, we define the array \gls{snr} as $\text{SNR}_{\text{A}} =  \mathbb{E}\{\lambda\|\bm{\Psi}[\bm{X}]_{t}\|^2\}$, $t=1,\dots,d$, where $[\bm{X}]_t$ denotes the $t$th column of $\bm{X}$ and the expectation does not depend on $t$ due to the assumed stationarity of the source amplitudes.

\begin{figure}
\centering
\setlength{\figurewidth}{0.82\columnwidth}
\setlength{\figureheight}{2.5cm}
\def\datapath{./pgf/}
\includegraphics{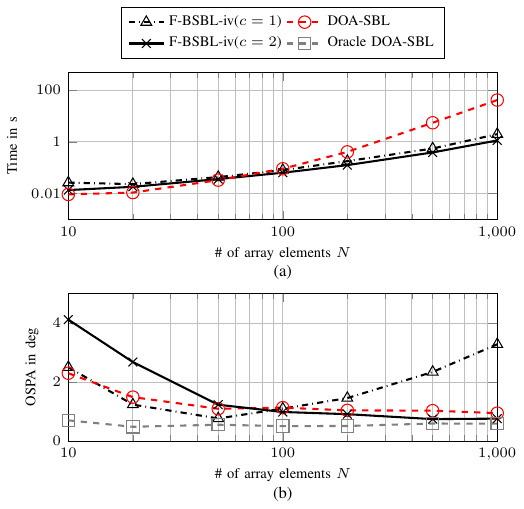}
\caption{Comparison of the runtime (a) and OSPA (b) of the \gls{fbsbl} algorithm and the DOA-SBL algorithm \cite{gerstoftSPL2016:SBLforDoA} versus the number of array elements at $\text{SNR}_{\text{A}}=10\,\text{dB}$.}
\label{fig:DOA:time}
\end{figure}

\subsubsection*{Estimation Performance in a Single Measurement Scenario}
For the first experiment, we analyze the runtime and estimation accuracy versus the number of array elements $N$ in such a scenario.
We generate a dictionary with $M=2N$ entries using a grid with samples spaced in such a way that its image through a $\sin$-transformation forms a regular grid of $(-1/2,+1/2]$. 
We simulate 3 sources located at the grid points closest to $\{-2^\circ,3^\circ,50^\circ\}$ with amplitudes drawn randomly from a zero-mean complex Gaussian distribution with unit variance.
We consider a single measurement scenario, i.e. $d=1$.
For the DOA-SBL algorithm, we calculate the \gls{ospa} based on all blocks with estimated component \gls{snr} $\geq 10\,\text{dB}$.
For the \gls{fbsbl} algorithm, we use again the hyperprior density $p(\gamma)=\gamma^{c-1}$ given in Row~iv of Table~\ref{tab:update-equations}.
To illustrate the effect of the parameter $c$, we use the \gls{fbsbl}-iv algorithm with two values of $c$, i.e. $c=1$ and $c=2$.
Figures \ref{fig:DOA:time}a and \ref{fig:DOA:time}b show the runtime and \gls{ospa}, respectively, for both algorithms at $\text{SNR}_{\text{A}}=20\,\text{dB}$.
The smallest \gls{ospa} is achieved by the oracle DOA-SBL algorithm. This is not surprising, since this algorithm is given the true number of sources in advance, which is not a realistic assumption for many practical applications.
For array sizes of 100 elements and larger, the \gls{fbsbl}-iv($c=2$) algorithm is faster than the DOA-SBL algorithm while achieving the same \gls{ospa}.
For array sizes with less than 100 elements, the \gls{ospa} of the \gls{fbsbl}-iv($c=2$) algorithm increases rapidly.
This is due to the rate of erroneously classifying active blocks as non-active, which rises when the number of array elements is decreased.
The parameter $c$ acts similarly to a threshold, see Section~\ref{sec:analysis:threshold}.
Hence, we can improve the block detection rate by using a smaller value for $c$.
Consequently, this also leads to an increased rate of erroneously classifying non-active blocks as active. This increases is mostly noticeable at large $N$, due to the larger number of \glspl{doa} points in the grid for large $N$.
Hence, the \gls{fbsbl}-iv($c=1)$ algorithm
reduces the \gls{ospa} at smaller array sizes at the cost of increasing the \gls{ospa} at larger array sizes compared to the \gls{fbsbl}-iv($c=2)$ algorithm, as shown in Figure \ref{fig:DOA:time}b.
We refer the reader to \cite{leitinger2020Asilomar} for a detailed discussion on the relation between the array size and the rate of erroneously classifying non-active blocks as active.

\subsubsection*{Estimation Performance in a Multiple Measurement Scenario}

\begin{figure}
\centering
\setlength{\figurewidth}{0.82\columnwidth}
\setlength{\figureheight}{2.5cm}
\def\datapath{./pgf/}
\includegraphics{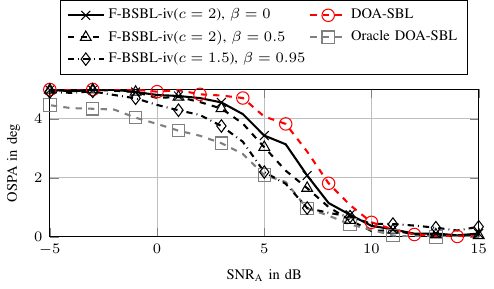}
\caption{Comparison of the OSPA of the \gls{fbsbl} algorithm and the DOA-SBL algorithm \cite{gerstoftSPL2016:SBLforDoA} versus $\text{SNR}_{\text{A}}$ with the source correlation parameter $\beta$ as a parameter in a MMV scenario with $d=10$ measurements.}
\label{fig:DOA:SNR}
\end{figure}

Next, we simulate a system with an array consisting of $N=100$ antennas from which $d=10$ measurements are obtained.
We simulate three sources at the same \glspl{doa} as in the previous experiment. To investigate the effect of intra-block correlation, the source amplitudes are generated by first-order \gls{ar} processes, i.e., for $i=1,\ldots,K$ $s_i[t] = \xi_i[t] + \beta s_i[t-1]$, $t=1,\ldots,d$, where all random variables $\xi_i[t]$ are independent, identically distributed, complex, Gaussian with zero mean and unit variance.
The coefficient $\beta\in \mathbb{C}$ : $|\beta| < 1$ sets the temporal correlation of the \gls{ar} processes.
The amplitudes of all sources have the same covariance matrix $\bm{\Sigma}_{\text{s}}$ equal to the Toeplitz matrix
\vspace*{-2mm}
\begin{equation}\label{eq:doa:ar-covariance}
\bm{\Sigma}_{\text{s}} = \left[\begin{matrix}
1 & \beta  & \cdots & \beta^{d-1} \\
\beta & 1 & \ddots & \beta^{d-2} \\
\vdots& \ddots & \ddots & \vdots \\
\beta^{d-1} & \beta^{d-2} & \cdots   & 1
\end{matrix}\right]
.
\vspace*{-2mm}
\end{equation}
The temporal correlation of the sources provides additional statistical information that can be exploited to separate true sources from additive white noise.
We evaluate three different cases: (i) no correlation $\beta=0$, (ii) medium correlation $\beta=0.5$ and (iii) strong correlation $\beta=0.95$.
For the \gls{fbsbl} algorithm we set $\bm{D}=\bm{\Sigma}_{\text{s}}^{-1}$ to exploit this information.
We use again the hyperprior density $p(\gamma)=\gamma^{c-1}$ given in Row~iv of Table~\ref{tab:update-equations}.
To achieve an (approximately) constant rate of erroneously classifying active blocks as non-active, we use \gls{fbsbl}-iv($c=2$) in the case of no correlation (i) and medium correlation (ii), and use \gls{fbsbl}-iv($c=1.5$) in the high correlation case (iii), based on preliminary simulations.
For the DOA-SBL algorithm, we calculate the \gls{ospa} based on all blocks with an estimated component \gls{snr} $\geq 2\,\text{dB}$, based on the same preliminary simulations.
The performance of the DOA-SBL algorithm is approximately the same in all three cases, since it is unable to exploit the information about the sources' temporal correlation.
Thus, we plot the performance of the DOA-SBL algorithm only for case (i).
Figure \ref{fig:DOA:SNR} depicts the \gls{ospa} of both algorithms versus the $\text{SNR}_{\text{A}}$.
For $\text{SNR}_{\text{A}} < 0\,\text{dB}$ and $\text{SNR}_{\text{A}} > 10\,\text{dB}$, the estimation either fails due to the high noise level or the points of the \gls{doa}-grid corresponding to the active sources are recovered with high probability by both algorithms. Thus, their performance is approximately the same in these regions.
In the transition region $0\,\text{dB} \leq \text{SNR}_{\text{A}} \leq 10\,\text{dB}$, all variants of the \gls{fbsbl} algorithm achieve a smaller \gls{ospa} than the DOA-SBL algorithm in all three cases. With increasing correlation $\beta$, the performance of the \gls{fbsbl} algorithm improves. For the case of high correlation (iii), the performance of the \gls{fbsbl}-iv($c=1.5$) algorithm is practically the same as the performance of the oracle DOA-SBL algorithm which is given the true number of sources $L$.
We refer the reader to \cite{zhang2011STSP:bSBL} for a more in-depth investigation and discussion of the effects of the intra-block correlation $\bm{D}$ as well as for suggestions on how to estimate this matrix efficiently without introducing too many additional parameters.

Note that in any practical example the sources' \glspl{doa} will not align exactly with the samples of the grid.
This model missmatch introduces additional errors in the estimation process.
These can be counteracted e.g., by using a variational-\gls{em} approach similar to \cite{shutin2013:VSBL,moederl2024TSP:structuredLSE} to optimize the \gls{elbo} over the estimated source \glspl{doa} in addition to the hyperparameters.
Furthermore, \cite{badiu2017TSP:VSBL} directly integrates the estimation of the \glspl{doa} into the \gls{vb} framework in order to obtain (approximate) posterior distributions of them.
However, we consider these off-grid approaches outside the scope of this work.

\section{Conclusion}

We present a variational implementation of \gls{bsbl}, coined \gls{vabsbl}, and derive a fast version of it, coined \gls{fbsbl}. The derivation of the latter makes use of a novel update rule derived from analyzing the fixed points of a first-order recurrence relation involving the hyperparameter mean estimates.
By analyzing the convergence behavior of this recurrence relation, we identify the range of the parameters of the generalized inverse Gaussian hyperprior for which the estimator inherently incorporates a pruning condition that switches off ``weak'' components in the model, which is necessary to obtain sparse results.

Numerical investigations demonstrate that \gls{fbsbl} achieves runtime improvements of up to two orders of magnitude compared to BSBL-EM and BSBL-BO \cite{zhang2013TSP:BlockSparseSBL}, while delivering superior \gls{nmse} across various signal conditions. Applied to \gls{doa} estimation, \gls{fbsbl}  outperforms the algorithm in \cite{gerstoftSPL2016:SBLforDoA} in terms of computation time in single measurement scenarios and achieves lower \gls{ospa} in low-\gls{snr}, correlated multiple measurement scenarios.

We show that \gls{em}- and \gls{vb}-based implementations of \gls{bsbl} are identical and thereby generalize an early result for classical \gls{sbl} to \gls{bsbl}.  As a consequence, the fast versions of these algorithms, namely \gls{bsbl} using coordinate ascent to maximize the marginal likelihood, e.g., \cite{ma2017TSP:FastBSBL}, and \gls{fbsbl} also coincide. 
The importance of this equivalence is underscored by the message-passing interpretation of the \gls{vb} implementation of \gls{bsbl}, which enables a principled merging of \gls{vb} and belief propagation \cite{riegler13}.

Promising extensions include estimation of the block sizes, application to models with continuous dictionary (e.g., in grid-free \gls{doa} estimation) akin to \cite{hansen2014SAM:SBL,hansen2018TSP:SuperFastLSE,moederl2024TSP:structuredLSE,GreLeiWitFle:TWC2024,shutin2013:VSBL}, and principled merging of \gls{fbsbl} with other message-passing methods for dedicated applications, such as joint channel estimation and decoding \cite{kirkelund10,hansenTSP2018:IterativeReceiverSBL}, sequential tracking of time-varying block-sparse channels \cite{xuhong2022TWC}, particularly in multiple-input\---multiple-output systems, for which the block-sparse assumption turns out to be realistic \cite{prasadTSP2015:MIMO-BlockSBL,barbu2016TVT:OFDM-BSBL}.

\appendix

\subsection{Proof of Lemma~\ref{lemma:expectation-as-polynomial-fraction}}
\label{sec:variational-fast-solution:update-polynomials}

We prove that the expectation $\big<\bm{x}_i^\hermitian \bm{D} \bm{x}_i\big>_{q_{\bm{x}}}$ in the update of $\hat{\gamma}_i$ can be written as a rational function $B_i(\hat{\gamma}_i)/A_i(\hat{\gamma}_i)$.
First, we find that
\begin{equation} \label{eq:expectation-over-weights}
\big<\bm{x}_i^\hermitian\bm{D}\bm{x}_i\big>_{q_{\bm{x}}} = \hat{\bm{x}}_i^\hermitian\bm{D}\hat{\bm{x}}_i + \text{tr}(\bm{D}\hat{\bm{\Sigma}}_i)
\end{equation}
where $\hat{\bm{x}}_i = \bm{E}_i^\transpose\hat{\bm{x}}$, $\hat{\bm{\Sigma}}_i=\bm{E}_i^\transpose\hat{\bm{\Sigma}}\bm{E}_i$, and $\bm{E}_i = [\bm{0} \iist \bm{I}_{d} \iist \bm{0}]^\transpose$ is an $M \times d$ selection matrix \cite[Eq. (378)]{petersen2012:matrixookbook}.
We use the Cholesky decomposition of $\bm{D}$, i.e., $\bm{L}_i\bm{L}_i^\hermitian = \bm{D}$, to rewrite the trace term in \eqref{eq:expectation-over-weights} as
	\footnote{For the case of $\bm{D}=\text{diag}(\bm{b}_i)$ being a diagonal matrix with the elements of the vector $\bm{b}_i$ on its main diagonal, the Cholesky decomposition results in the matrix $\bm{L}_i=\text{diag}(\sqrt{\bm{b}_i})$.
    Similarly, if $\bm{D}=\bm{I}$ then it follows that $\bm{L}_i=\bm{I}$. Subsequently, $\bm{L}_i$ can be removed from the definition of $\bm{q}_i = \hat{\lambda}\bm{U}_i^\hermitian\bm{E}_i^\transpose \hat{\bm{\Sigma}}_{\sim i}\bm{\Phi}^\hermitian\bm{y}$, and $s_{i,l}$, $l=1,\dots,d$, are the eigenvalues of $\bm{E}_i^\transpose\hat{\bm{\Sigma}}_{\sim i}\bm{E}_i$.}
\begin{equation}\label{eq:trace-term}
\text{tr}(\bm{D}\hat{\bm{\Sigma}}_i) = \text{tr}(\bm{L}_i^\hermitian\bm{E}_i^\transpose\hat{\bm{\Sigma}}\bm{E}_i\bm{L}_i)
\end{equation}
and make the dependence on $\hat{\gamma}_i$ explicit by writing $\hat{\bm{\Sigma}} = \big(\hat{\bm{\Sigma}}_{\sim i}^{-1} + \hat{\gamma}_{i}\bm{E}_i\bm{D}\bm{E}_i^\transpose \big)^{-1}$ and applying the Woodbury matrix identity to obtain
\begin{equation}
\hat{\bm{\Sigma}} = \hat{\bm{\Sigma}}_{\sim i} - \hat{\bm{\Sigma}}_{\sim i} \bm{E}_i(\hat{\gamma}_i^{-1}\bm{D}^{-1} + \bm{E}_i^\transpose\hat{\bm{\Sigma}}_{\sim i}\bm{E}_i)^{-1}\bm{E}_i^\transpose\hat{\bm{\Sigma}}_{\sim i}
\label{eq:sigma-woodbury-identity}
\end{equation}
where $\hat{\bm{\Sigma}}_{\sim i}=\big(\lambda \bm{\Phi}^\hermitian\bm{\Phi} + \sum_{k=1,\,k\neq i}^K 	\hat{\gamma}_{k}\bm{E}_k\bm{D}\bm{E}_k^\transpose \big)^{-1}$.
Let $\bm{L}_i^\hermitian\bm{E}_i^\transpose\hat{\bm{\Sigma}}_{\sim i}\bm{E}_i\bm{L}_i=\bm{U}_i\bm{S}_i\bm{U}_i^\hermitian$ be the eigendecomposition of $\bm{L}_i^\hermitian\bm{E}_i^\transpose\hat{\bm{\Sigma}}_{\sim i}\bm{E}_i\bm{L}_i$, i.e., its eigenvectors are the columns of the unitary matrix $\bm{U}_i$ and its eigenvalues $s_{i,l}$, $l=1,\ldots,d$ are the diagonal elements of the diagonal matrix $\bm{S}_i$.
We insert \eqref{eq:sigma-woodbury-identity} into \eqref{eq:trace-term} and use the identity
\begin{equation} \label{eq:matrix-inverse-identity}
(\hat{\gamma}_i^{-1}\bm{D}^{-1} + \bm{E}_i^\transpose\hat{\bm{\Sigma}}_{\sim i}\bm{E}_i)^{-1} = \bm{L}_i\bm{U}_i(\hat{\gamma}_i^{-1}\bm{I}_{d} + \bm{S}_i)^{-1}\bm{U}_i^\hermitian\bm{L}_i^\hermitian
\end{equation}
to rewrite the trace term in \eqref{eq:expectation-over-weights} as
\begin{align} \label{eq:gamma-update-trace-term}
\text{tr}\big(\bm{D}\hat{\bm{\Sigma}}_i\big)
&=\text{tr}\big(\bm{S}_i\big) - \text{tr}\big(\bm{S}_i(\hat{\gamma}_i^{-1}\bm{I}_{d} + \bm{S}_i)^{-1}\bm{S}_i \big) \nn \\
&= \sum_{l=1}^{d} \frac{s_{i,l}}{1+\hat{\gamma}_i s_{i,l}}
\, .
\end{align}

Next, we investigate the expression $\hat{\bm{x}}_i^\hermitian\bm{D}\hat{\bm{x}}_i$ in \eqref{eq:expectation-over-weights}.
Using \eqref{eq:weights-mean}, \eqref{eq:sigma-woodbury-identity} and \eqref{eq:matrix-inverse-identity}, we find 
\begin{align}
\hat{\bm{x}}_i^\hermitian\bm{D}\hat{\bm{x}}_i &=
\hat{\lambda}^2 \bm{y}^\hermitian\bm{\Phi}\hat{\bm{\Sigma}}_{\sim i}\bm{E}_i\bm{L}_i\bm{U}_i[\bm{I}_{d} - 2 \bm{S}_i(\hat{\gamma}_i^{-1}\bm{I}_{d}+\bm{S}_i)^{-1} \nn \\
&\quad +\bm{S}_i^2(\hat{\gamma}_i^{-1}\bm{I}_{d}+\bm{S}_i)^{-2}]\bm{U}_i^\hermitian\bm{L}_i^{\hermitian}\bm{E}_i^\transpose\hat{\bm{\Sigma}}_{\sim i}\bm{\Phi}^\hermitian\bm{y}
\label{eq:gamma-update-amplitude-term}
\end{align}
after a few algebraic manipulations.
Equation \eqref{eq:gamma-update-amplitude-term} is a quadratic form $\bm{q}_i^\hermitian\bm{\Lambda}_i\bm{q}_i$ of the vector 
\begin{equation} \label{eq:definition-qi}
\bm{q}_i=\hat{\lambda}\bm{U}_i^\hermitian\bm{L}_i^{\hermitian}\bm{E}_i^\transpose\hat{\bm{\Sigma}}_{\sim i}\bm{\Phi}^\hermitian\bm{y}
\end{equation}
with the diagonal matrix $\bm{\Lambda}_i = \bm{I}_{d} - 2\bm{S}_i(\hat{\gamma}_i^{-1}\bm{I}_{d} + \bm{S}_i)^{-1} + \bm{S}_i^2(\gamma_i^{-1}\bm{I}_{d} +\bm{S}_i)^{-2}$, which can be expressed as
\begin{align}	
\hat{\bm{x}}_i^\hermitian\bm{D}\hat{\bm{x}}_i
&=\sum_{l=1}^{d}\frac{|q_{i,l}|^2}{(1+\hat{\gamma}_i s_{i,l})^2}
\label{eq:update-sum-2}
\end{align}
where $q_{i,l}$ is the $l$-th entry of the vector $\bm{q}_i$.

We define the polynomial of degree $2 d$
\begin{align}
A_i(\gamma) = \prod_{l=1}^{d} (1+\gamma s_{i,l})^2
\label{eq:polynom-A}
\end{align}
and the polynomial of degree $2 d-1$
\begin{align}
B_i(\gamma) = \sum_{l=1}^{d}(\gamma s_{i,l}^2 + |q_{i,l}|^2 + s_{i,l})\prod_{j=1, j\neq i}^{d} (1+\gamma s_{i,j})^2
.
\, \nn \\[-5mm]
\label{eq:polynom-B}
\end{align}
Inserting \eqref{eq:gamma-update-trace-term} and \eqref{eq:update-sum-2} into \eqref{eq:expectation-over-weights}, we arrive at $\big<\bm{x}_i^\hermitian\bm{D}\bm{x}_i\big>_{q_{\bm{x}}}=B_i(\hat{\gamma}_i)/A_i(\hat{\gamma}_i)$ after a few algebraic manipulations.

\subsection{Proof of Lemma~\ref{lemma:decreasing-expectation}}
\label{sec:appendix:derivative-hi}
We recast $B_i(\gamma)$ given in \eqref{eq:polynom-B} as
\begin{align}\label{eq:appendix:Polynomial-B-rewrite}
B_i(\gamma) 
&=\sum_{l=1}^{d} \frac{\gamma s_{i,l}^2 + |q_{i,l}|^2 + s_{i,l}}{(1+\gamma s_{i,l})^2}\prod_{j=1}^{d} (1+\gamma s_{i,j})^2 
\,.
\end{align}
Using \eqref{eq:appendix:Polynomial-B-rewrite}, we obtain
\begin{equation}
\label{eq:hi-sum}
\frac{B_i(\gamma)}{A_i(\gamma)} = \sum_{l=1}^{d} \frac{\gamma s_{i,l}^2 + |q_{i,l}|^2 + s_{i,l}}{(1+\gamma s_{i,l})^2}
\, .
\end{equation}
Taking the derivative of \eqref{eq:hi-sum} with respect to $\gamma$ yields
\begin{align}
\bigg(\frac{B_i(\gamma)}{A_i(\gamma)}\bigg)^\prime
&= - \sum_{l=1}^{d} \frac{\gamma s_{i,l}^3 + 2s_{i,l}|q_{i,l}|^2 + s_{i,l}^2}{(1+\gamma s_{i,l})^3}
\,.
\end{align}
where $s_{i,l}>0$, $i=1,\dots,K$, $l=1,\dots,d$, since they are eigenvalues of a positive definite matrix.
Hence, the derivative is negative, i.e., $B_i(\gamma) / A_i(\gamma)$ is decreasing.

\subsection{Proof of Lemma~\ref{lemma:besselK-ratio}}
\label{sec:appendix:fi-increasing}

We prove that
\begin{align}
h_{\alpha}(u) = \frac{1}{\sqrt{u}}\frac{K_{\alpha+1}(\sqrt{u})}{K_{\alpha}(\sqrt{u})}
\end{align}
is strictly decreasing by showing that
\begin{align}
\label{eq:proof-fi-1}
h'_{\alpha}(u) < 0
.
\end{align}
Starting from \cite[Eq. (9.6.26)]{abramowitz1972:handbookMathFun}
\begin{align}
K'_{\alpha}(z) = - K_{\alpha+1}(z)+\frac{\alpha}{z}K_{\alpha}(z)
\end{align}
a change of variables to $z=\sqrt{u}$ yields after some algebraic manipulations
\begin{align}
\frac{1}{\sqrt{u}}\frac{K_{\alpha+1}(\sqrt{u})}{K_{\alpha}(\sqrt{u})} = -2\bigg[\frac{
K'_{\alpha}(\sqrt{u})}{K_{\alpha}(\sqrt{u})} - \frac{\alpha/2}{u}\bigg]
.
\end{align}
Inserting $\frac{
K'_{\alpha}(\sqrt{u})}{K_{\alpha}(\sqrt{u})}=\frac{\mathrm{d}}{\mathrm{d}u}\ln K_{\alpha}(\sqrt{u})$ and $-\frac{\alpha/2}{u}=\frac{\mathrm{d}}{\mathrm{d}u}\ln u^{-\alpha/2}$, we find
\begin{align}
\label{eq:proof-fi-2}
h_{\alpha}(u) &= \frac{\mathrm{d}}{\mathrm{d}u} \Big[ -2 \ln \big(u^{-\alpha/2}K_{\alpha}(\sqrt{u})\big)\Big] 
=
k'(u)
\end{align}
where we defined $k(u) = -2\ln \big( u^{-\alpha/2} K_{\alpha}(\sqrt{u})\big)$.

From \eqref{eq:proof-fi-2} follows that the condition \eqref{eq:proof-fi-1} is equivalent to 
\begin{align}
k''(u)<0\ist,    
\end{align}
i.e., that $h_{\alpha}(u)$ is strictly decreasing if, and only if, $k(u)$ is strictly concave.
We write $k(u)=k_1\circ k_2(u)$ as the composition of $k_1(v)=-2\ln v$ and $k_2(u) = u^{-c/2}K_{\alpha}(\sqrt{u})$ to obtain
\begin{align}
k^{\prime\prime}(u) = k_1^{\prime\prime}(k_2(u)) \cdot k_2^\prime(u)^2 + k_1^\prime(k_2(u)) \cdot k_2^{\prime\prime}(u)\ist.
\end{align}
Inserting $k_1^\prime(v) = -\frac{2}{v}$ and $k_1^{\prime\prime}(v)=\frac{2}{v^2}$ we find
\begin{align}
\label{eq:proof-fi-5}
k^{\prime\prime}(u) = 2 \frac{k_2^{\prime}(u)}{k_2(u)}\Big[\frac{k_2^{\prime}(u)}{k_2(u)} - \frac{k_2^{\prime\prime}(u)}{k_2^\prime(u)} \Big]
.
\end{align}
Using \cite[Eq. (9.6.28)]{abramowitz1972:handbookMathFun}, we express the derivatives $k_2^\prime(u)$ and $k_2^{\prime\prime}(u)$ as
\begin{align}
k_2^{\prime}(u) &= -\frac{1}{2} u^{-(\alpha+1)/2} K_{\alpha+1}(\sqrt{u}) \\
k_2^{\prime\prime}(u) &= \frac{1}{4} u^{-(\alpha+2)/2}K_{\alpha+2}(\sqrt{u})
\end{align}
to obtain
\begin{align}
\label{eq:proof-fi-3}
\frac{k_2^\prime(u)}{k_2(u)} &= -\frac{1}{2}u^{-1/2}R_{\alpha}(\sqrt{u}) \\
\frac{k_2^{\prime\prime}(u)}{k_2^\prime (u)} &= -\frac{1}{2}u^{-1/2}R_{\alpha+1}(\sqrt{u})
\label{eq:proof-fi-4}
\end{align}
where $R_{\alpha}(z)=\frac{K_{\alpha+1}(z)}{K_{\alpha}(z)}$.
For $z>0$, the function $R_{\alpha}(z)$ is positive and increasing with respect to $\alpha$ \cite[Lemma 2.2]{ismail1978}.
Thus, inserting \eqref{eq:proof-fi-3} and \eqref{eq:proof-fi-4} into \eqref{eq:proof-fi-5}, we arrive at
\begin{align}
\hspace*{-1ex}
k''(u) &= -\frac{1}{2}u^{-1}R_{\alpha}(\sqrt{u}) \big[R_{\alpha+1}(\sqrt{u}) - R_{\alpha}(\sqrt{u})\big] < 0.
\end{align}

\subsection{Conditions for the Convex Representation}
\label{sec:appendix:conditions-convex-representation}

We show that the \gls{pdf} $p(\tilde{\bm{x}})$ in \eqref{eq:ghd}  admits the convex representation \eqref{eq:prior-convex-type}. 
Defining $\alpha=-\hat{c} = -(c+\rho d)$, we recast \eqref{eq:ghd} as
\begin{align}
p(\tilde{\bm{x}}) \propto \Big(\sqrt{b(a+z^2)}\Big)^{\alpha} K_{\alpha}\Big(\sqrt{b(a+z^2)}\Big) 
\end{align}
with $z=\sqrt{\tilde{\bm{x}}^\hermitian\bm{D}\tilde{\bm{x}}}$.
We rewrite the right-hand side function of $z$ as $\exp\big(-g(z^2)\big)$ with
\begin{align}
g(z) &= -\frac{\alpha}{2} \ln\big(b(a+z)\big) - \ln K_{\alpha}\big(\sqrt{b(a+z)}\big) + \text{const.} \nn \\[-2mm]
\end{align}
We show now that $g(z)$ is increasing and concave. Calculating the derivative of $g(z)$, we find
\begin{align}\label{eq:convex-conditions-first-derivative}
	g'(z) = \frac{b}{2}\Big(\sqrt{b(a+z)}R_{\alpha-1}\big(\sqrt{b(a+z)}\big)\Big)^{-1}
\end{align}
using \cite[Eq. (9.6.26)]{abramowitz1972:handbookMathFun} and $R_{\alpha-1}(u)= K_{\alpha}(u)/K_{\alpha-1}(u)$.
Thus, $g(z)$ is increasing since $R_{\alpha}$ is positive on $(0,\infty)$ \cite{jorgensen1982}.

Writing \eqref{eq:convex-conditions-first-derivative} as the composition
\begin{align}
g'(z) = \frac{b}{2}r_{\alpha}^{-1} \circ u(z)
\end{align}
of the two functions $u(z)=\sqrt{b(a+z)}$ and $r_{\alpha}(u) = u R_{\alpha-1}(u)$, we obtain the second derivative as
\begin{align}
g''(z) &= -\frac{b}{2}r_{\alpha}(u(z))^{-2} \cdot r_{\alpha}^{\prime}(u(z)) \cdot u^\prime(z)
\end{align}
where $r_\alpha^\prime$ and $u^\prime$ are the first derivatives of $r_{\alpha}$ and $u$, respectively. The function $r_{\alpha}$ is increasing for any real $\alpha$, i.e., $r_{\alpha}^\prime(u) > 0$ \cite[Lemma~2.6]{ismail1978}. Thus, $g(z)$ is concave since $r_{\alpha}(u)>0$ and $u^\prime(z)>0$ also hold for any $u,z\in (0,\infty)$.

\subsection{Detailed Derivations for Theorem~\ref{theorem:em-equivalent}}
\label{sec:appendix:em-vb-equivalence}

First, we investigate the \gls{em} algorithm based on the convex representation \eqref{eq:prior-convex-type} of the \gls{pdf} $p(\bm{x}_i)$.
Inserting $p(\bm{x}_i;\gamma_i)=\mathrm{N}\big(\bm{x}_i;\,\bm{0},\,(\gamma\bm{D})^{-1}\big)$ into \eqref{eq:prior-convex-type}, we find after some algebraic manipulations
\begin{align} \label{eq:em-vb-equivalence-1}
-\ln p(\bm{x}_i) = -\rho \ln \Big|\frac{\rho}{\pi}\bm{D}\Big| + \omega_i(v_i)\big|_{v_i=\rho \bm{x}_i^\hermitian\bm{D}\bm{x}_i},
\end{align}
where 
\begin{align} \label{eq:em-vb-equivalence-2}
\omega_i(v_i) = \inf_{\gamma_i} \big\{ v_i \gamma_i - \rho d \ln \gamma_i - \ln \varphi(\gamma_i) \big\}
\end{align}
is a concave function of $v_i\in\mathbb{R}_{+}$ with dual $\omega_i^\ast(\gamma_i)=- \rho d \ln \gamma_i - \ln \varphi(\gamma_i)$.

We insert $q_{\bm{x}} = \mathrm{N}(\bm{x};\,\hat{\bm{x}},\hat{\bm{\Sigma}})$, with $\hat{\bm{x}}$ and $\hat{\bm{\Sigma}}$ given by \eqref{eq:weights-mean} into \eqref{eq:em-Q-function}, to find
\begin{align}
\gamma_{i}^{\mathrm{EM}} &= \arg\max_{\gamma_i} Q\big(\bm{\gamma},q_{\bm{x}}\big)
\nn \\
&= \arg\min_{\gamma_i} \big\{ \gamma_i \,\rho\big<\bm{x}_i^\hermitian \bm{D}\bm{x}_i\big>_{q_{\bm{x}}} - \rho d \ln \gamma_i - \ln \varphi(\gamma_i) \big\} \nn \\
&= \arg\min_{\gamma_i} \{v_i \gamma_i - \omega_i^\ast(\gamma_i) \}\big|_{v_i=\rho \langle\bm{x}_i^\hermitian\bm{D}\bm{x}_i\rangle_{q_{\bm{x}}}}
\, .
\label{eq:em-vb-equivalence-3}
\end{align}
To find the minimum point of the function given in the braces in \eqref{eq:em-vb-equivalence-3}, we set its first derivative to zero, i.e., $\big(v_i \gamma_i - \omega_i^\ast(\gamma_i)\big)^\prime=0$, resulting in
\begin{align} \label{eq:em-vb-equivalence-4}
\gamma_{i}^{\mathrm{EM}} = \omega_i^\prime(v_i)\big|_{v_i
= \rho \langle\bm{x}_i^\hermitian\bm{D}\bm{x}_i\rangle_{q_{\bm{x}}}}
\, .
\end{align}

Next, we turn to the \gls{vb} algorithm. Let
\begin{align}\label{eq:em-vb-equivalence-7}
q_{\gamma_i; v_i}(\gamma_i) 
\propto \gamma_i^{\rho d} \exp \{ -v_i \gamma_i\} \cdot p(\gamma_i)
\,
\end{align}
denote a \gls{pdf} parameterized by $v_i\in\mathbb{R}_{+}$ the \glspl{pdf}
\begin{align}\label{eq:em-vb-equivalence-6}
q^\star_{\gamma_i}(\gamma_i)
\propto \gamma^{\rho d} \exp\{ -\rho \gamma_i \big<\bm{x}_i^\hermitian \bm{D} \bm{x}_i\big>_{q_{\bm{x}}} \} \cdot p(\gamma_i)
\end{align}
obtained from \eqref{eq:precision-distribution} and
\begin{equation}
    p(\gamma_i|\bm{x}_i)\propto \gamma^{\rho d}\exp\{-\rho \gamma_i \bm{x}_i^\hermitian\bm{D}\bm{x}_i\}\cdot p(\gamma_i)
\end{equation}
can be both expressed as \eqref{eq:em-vb-equivalence-7} with $v_i=\rho \big<\bm{x}_i^\hermitian \bm{D} \bm{x}_i\big>_{q_{\bm{x}}}$ and $v_i=\rho \bm{x}_i^\hermitian \bm{D} \bm{x}_i$, respectively.

Combining \eqref{eq:em-vb-equivalence-6} and \eqref{eq:em-vb-equivalence-7}, we write the \gls{vb} update as
\begin{equation}\label{eq:em-vb-equivalence-14}
    \hat{\gamma}_i = \big<\gamma_i\big>_{q^\star_{\gamma_i}} = \big<\gamma_i\big>_{q_{\gamma_i;v_i}}\big|_{v_i=\rho \langle\bm{x}_i^\hermitian\bm{D}\bm{x}_i\rangle_{q_{\bm{x}}}}
\end{equation}
which is equivalent to the \gls{em} update \eqref{eq:em-vb-equivalence-4} if
\begin{equation} \label{eq:em-vb-equivalence-13}
    \big<\gamma_i\big>_{q_{\gamma_i;v_i}}=\omega_i^\prime(v_i)\quad,\quad v_i\in\mathbb{R}_{+}
\end{equation}
because the proxy \gls{pdf} $q_{\bm{x}}$ used in the \gls{em} equation \eqref{eq:em-vb-equivalence-4} and the \gls{vb} equation \eqref{eq:em-vb-equivalence-14} are identical, i.e., $q_{\bm{x}}=q^\star_{\bm{x}}=q^\text{EM}_{\bm{x}}$, as can be easily checked from \eqref{eq:weight-proxypdf} and \eqref{eq:em-Q-function} when $\hat{\bm{\gamma}}$ in the former equation and $\bm{\gamma}$ in the latter are set equal.
To show that \eqref{eq:em-vb-equivalence-13} holds, 
we first introduce some intermediate results.
Using the rules of differentiating under the integral and standard vector calculus, it can be shown that
the gradient of $p(\bm{x}_i|\gamma_i)$ as a function of $\bm{x}_i$ is given by
\begin{align}\label{eq:em-vb-equivalence-8}
\nabla p(\bm{x}_i|\gamma_i) = -\gamma_i \bm{D}\bm{x}_i p(\bm{x}_i|\gamma_i)\ist .
\end{align}
Using the identity $p(\bm{x}_i) = \int p(\bm{x}_i|\gamma_i)p(\gamma_i)\,\mathrm{d}\gamma_i$, we find
\begin{align}
\nabla p(\bm{x}_i)
&= -\bm{D} \bm{x}_i p(\bm{x}_i) \cdot
\big<\gamma_i\big>_{p(\gamma_i|\bm{x}_i)}
\, .
\label{eq:em-vb-equivalence-9}
\end{align}
Rewriting \eqref{eq:em-vb-equivalence-1} as the composition
\begin{align}
p(\bm{x}_i) = \big(\frac{\rho}{\pi}\big)^{\rho d} |\bm{D}|^\rho \exp \{ -\omega_i(v_i)\} \circ v_i(\bm{x}_i)
\end{align} 
with $v_i(\bm{x}_i)=\rho\bm{x}_i^\hermitian \bm{D} \bm{x}_i$ we calculate the gradient to be
\begin{align} \label{eq:em-vb-equivalence-10}
\nabla p(\bm{x}_i) = -\bm{D} \bm{x}_i p(\bm{x}_i)  \cdot \omega_i^\prime(v_i)|_{v_i=\rho \bm{x}_i^\hermitian \bm{D} \bm{x}_i} 
\, .
\end{align}
Combining \eqref{eq:em-vb-equivalence-9} with \eqref{eq:em-vb-equivalence-10}, we find that
\begin{align}\label{eq:em-vb-equivalence-11}
\omega_i^\prime\big(v_i)|_{v_i=\rho \bm{x}_i^\hermitian\bm{D}\bm{x}} &= \big<\gamma_i\big>_{p(\gamma_i|\bm{x}_i)}
= \big<\gamma_i\big>_{q_{\gamma_i;\,v_i}}\big|_{v_i=\rho \bm{x}_i^\hermitian\bm{D}\bm{x}_i}
\, .
\end{align}
Since $\bm{D}$ is positive definite and \eqref{eq:em-vb-equivalence-11} holds for any $\bm{x}_i\in\mathbb{\mathbb{R}}^d$ or $\bm{x}_i \in \mathbb{C}^{d}$ (i.e., any $v_i\in\mathbb{R}_{+}$), \eqref{eq:em-vb-equivalence-13} is true.

\subsection{Equivalence of F-BSBL and the Coordinate-Ascent Implementation of BSBL}
\label{sec:appendix:equivalence-prove}

We show that the coordinate-ascent implementation of \gls{bsbl}, e.g., \cite{ma2017TSP:FastBSBL}, is identical to \gls{fbsbl}. For ease of notation, we assume the noise precision $\lambda$ fixed and known.
We consider $\varphi(\gamma)=1$. We obtain an analytic expression for the marginal likelihood by solving the integral in \eqref{eq:em-lower-bound}, yielding
\begin{equation} \label{eq:marginal-likelihood-full}
\tilde{p}(\bm{y};\bm{\gamma}) = \ln | \bm{\Sigma}| + \lambda^2\bm{y}^\hermitian\bm{\Phi}\bm{\Sigma}\bm{\Phi}^\hermitian\bm{y} + \sum_{i=1}^{K} d \ln \gamma_i + \text{const.}
\end{equation}
where $\bm{\Sigma}=(\lambda\bm{\Phi}^\hermitian\bm{\Phi} + \text{diag}(\bm{\gamma})\otimes \bm{D})^{-1}$.
First, we show that the dependency of the marginal likelihood on a single entry $\gamma_i\mapsto \tilde{p}(\bm{y};\bm{\gamma})$
can be expressed as $\tilde{p}(\bm{y};\bm{\gamma})=\ell_i(\gamma_i) + \text{const.}$, where
\begin{align}\label{eq:marginal-likelihood-singleParam}
\ell_i(\gamma_i) &= \sum_{l=1}^{d} \ln \bigg(\frac{\gamma_i s_{i,l}}{1+\gamma_i s_{i,l}}\bigg) - \frac{\gamma_i |q_{i,l}|^2}{1+\gamma_i s_{i,l}}
\end{align}
with $s_{i,l}$ and $q_{i,l}$ defined in Appendix~\ref{sec:variational-fast-solution:update-polynomials}.

Let $\bm{\Phi}_i$ be the matrix consisting of all columns of the dictionary matrix corresponding to the $i$th block and $\bm{\Phi}_{\sim i}$ the matrix containing all remaining columns of $\bm{\Phi}$ and rearrange $\bm{\Phi}$, such that $\bm{\Phi} = [\bm{\Phi}_{\sim i}\iist \bm{\Phi}_i]$.
Finally, let us define $\bm{\Gamma}_{\sim i}=\text{diag}(\bm{\gamma}_{\sim i})\otimes \bm{D}$ where $\bm{\gamma}_{\sim i}$ is the vector $\bm{\gamma}$ with the $i$th element removed. We write $\bm{\Sigma}$ as a block matrix
\begin{align} \label{eq:equivalence-marginal-likelihood-1}
\bm{\Sigma} &= \left[\begin{matrix}
\lambda\bm{\Phi}_{\sim i}^\hermitian\bm{\Phi}_{\sim i} + \bm{\Gamma}_{\sim i} & \lambda \bm{\Phi}_{\sim i}^\hermitian\bm{\Phi}_i \\
\lambda \bm{\Phi}_i^\hermitian\bm{\Phi}_{\sim i} & \lambda \bm{\Phi}_i^\hermitian\bm{\Phi}_i + \gamma_i\bm{D}
\end{matrix}\right]^{-1} 
\end{align}
and apply the block determinant lemma to the first term in \eqref{eq:marginal-likelihood-full} with the right-hand term  in \eqref{eq:equivalence-marginal-likelihood-1} to obtain
\begin{align} \label{eq:fastMarginal-equivalence-log-term}
\ln|\bm{\Sigma}|
&= \text{const.} - \sum_{l=1}^{d} \ln \big(\gamma_i+ \frac{1}{s_{i,l}}\big)
\, .
\end{align}
To make the dependence on $\gamma_i$ in the second term in \eqref{eq:marginal-likelihood-full} explicit, we recast $\bm{\Sigma} = \big(\bm{\Sigma}_{\sim i}^{-1} + \gamma_i\bm{E}_i\bm{D}\bm{E}_i^\transpose\big)^{-1}$, where $\bm{\Sigma}_{\sim i}=\big(\lambda \bm{\Phi}^\hermitian\bm{\Phi} + \sum_{k=1,\,k\neq i}^K \gamma_{k}\bm{E}_k\bm{D}\bm{E}_k^\transpose \big)^{-1}$, and use the Woodbury matrix identity together with \eqref{eq:matrix-inverse-identity}, which yields
\begin{align}\label{eq:fastMarignal-equivalence-data-term-2}
\lambda^2\bm{y}^\hermitian\bm{\Phi}\bm{\Sigma}\bm{\Phi}^\hermitian\bm{y} &= \text{const.} -\bm{q}_i^\hermitian(\gamma_i^{-1}\bm{I} +\bm{S}_i)^{-1}\bm{q}_i \nn \\
&= \text{const.} - \sum_{l=1}^{d}\frac{\gamma_i |q_{i,l}|^2}{1 + \gamma_i s_{i,l}}
\end{align}
where $\bm{q}_i$ is the vector defined in \eqref{eq:definition-qi}.
Inserting \eqref{eq:fastMarginal-equivalence-log-term} and \eqref{eq:fastMarignal-equivalence-data-term-2} into \eqref{eq:marginal-likelihood-full} we arrive at \eqref{eq:marginal-likelihood-singleParam}.

To find the maximum of $\gamma_i\mapsto \tilde{p}(\bm{y};\bm{\gamma})$ we set the partial derivative of \eqref{eq:marginal-likelihood-singleParam} with respect to $\gamma_i$ to zero to obtain the fixed-point equation
\begin{equation} \label{eq:marginal-likelihood-derivative}
\ell_i^\prime(\gamma_i) = \sum_{l=1}^{d}\frac{1-\gamma(|q_{i,l}|^2-s_{i,l})}{\gamma_i(1+\gamma_i s_{i,l})^2} = 0
\end{equation}
which can be expressed as
\begin{align} 
0 
&= d A_i(\gamma_i) -  \gamma_i B_i(\gamma_i)
\label{eq:marginal-likelihood-solution}
\end{align}
after some algebraic manipulations.
In \eqref{eq:marginal-likelihood-solution}, $A_i(\gamma_i)$ and $B_i(\gamma)$ are the polynomials defined in \eqref{eq:polynom-A} and \eqref{eq:polynom-B}, respectively.
The solutions of \eqref{eq:marginal-likelihood-solution} are, by definition, the roots of the polynomial $G_i(\gamma)$ given in Row~v of Table~\ref{tab:update-equations}.
Hence, the extrema of $\gamma_i\mapsto \tilde{p}(\bm{y};\bm{\gamma})$ correspond to the fixed points of the recurrent relation in the fast \gls{vb} implementation of \gls{bsbl} in case Jeffrey's prior is used.

\bibliography{IEEEabrv,fastBlockSBL_references}
\bibliographystyle{IEEEtran}

\end{document}